\documentclass[a4paper,fleqn]{cas-sc}

\usepackage[numbers]{natbib}
\usepackage{amsmath,amssymb,bm,booktabs}
\usepackage[caption=false]{subfig}
\usepackage{cleveref}
\usepackage{algorithm}
\usepackage{algpseudocode}
\usepackage{tikz}
\usetikzlibrary{arrows.meta,calc,positioning,shapes.geometric}

\newcommand{\dd}{\mathrm{d}}

\newdefinition{remark}{Remark}
\newproof{proof}{Proof}

\crefname{proposition}{proposition}{propositions}
\Crefname{proposition}{Proposition}{Propositions}
\crefname{lemma}{lemma}{lemmas}
\Crefname{lemma}{Lemma}{Lemmas}
\crefname{remark}{remark}{remarks}
\Crefname{remark}{Remark}{Remarks}

\begin{document}
\let\WriteBookmarks\relax
\def\floatpagepagefraction{1}
\def\textpagefraction{.001}

\shorttitle{GSIS for the non-gray phonon BTE}
\shortauthors{D. Shen, J. Liu and W. Su}

\title[mode=title]{A Synthetic Iterative Scheme for Non-Gray Phonon Boltzmann Transport Equation with Dual Relaxation Times}

\author[1]{Dingtao Shen}
\ead{dshenad@connect.ust.hk}

\affiliation[1]{organization={Division of Emerging Interdisciplinary Areas, Academy of Interdisciplinary Studies, The Hong Kong University of Science and Technology},
            city={Hong Kong SAR},
            country={China}}

\author[1]{Jia Liu}
\ead{jliugh@connect.ust.hk}

\author[1,2]{Wei Su}
\cormark[1]
\ead{weisu@ust.hk}

\affiliation[2]{organization={Department of Mathematics, The Hong Kong University of Science and Technology},
            city={Hong Kong SAR},
            country={China}}

\cortext[1]{Corresponding author}

\begin{abstract}
Solving the non-gray Callaway phonon Boltzmann transport equation allows a dual-relaxation-time approximation of separate normal and resistive scatterings and resolving the mode-dependent spectrum. The conventional iterative scheme (CIS) for deterministic solutions avoids a monolithic phase-space inversion, but its collision-source iteration can become prohibitively slow at a large characteristic length of a material. Existing synthetic acceleration schemes address either non-gray single-relaxation models or gray dual-relaxation models, leaving mode-resolved dual-relaxation transport without a dedicated acceleration framework. We develop a general synthetic iterative scheme (GSIS) for the stationary, linearized, non-gray Callaway equation, where synthetic approximations for the normal-process pseudo-temperature and phonon drift velocity are provided by exact energy and quasi-momentum balance laws closed with first-order Chapman-Enskog constitutive relations and non-equilibrium terms evaluated from the kinetic solution. The resistive-process pseudo-temperature is retrieved from the two quantities. Each iteration couples an upwind nodal discontinuous Galerkin kinetic sweep and a hybridizable discontinuous Galerkin solution of the synthetic equations to achieve high-order spatial discretization. A branch- and frequency-resolved Fourier analysis identifies the deterioration of CIS and shows that the GSIS contraction factor remains bounded away from unity for the considered graphene material. Asymptotic analysis indicates that GSIS reduces to a consistent discretization of a Guyer-Krumhansl-like equation and Fourier's law of heat conduction in the hydrodynamic and diffusive limits, respectively.

Numerical experiments reproduce established non-gray graphene benchmarks and confirm that CIS and GSIS converged the same discrete solution. Across fifty-six jointly converged benchmark points, GSIS achieves maximum speed-up factors in several tens for both iteration steps and wall-clock time. Tests performed on unstructured meshes, variant boundary conditions, and an extended size sweep from 1 nm to 1 mm further show that GSIS remains converged within up to 4000 iterations. At the same time, CIS reaches the prescribed iteration cap of $2\times10^5$ after 100 $\mu$m.

\end{abstract}

\begin{keywords}
Phonon Boltzmann equation \sep General synthetic iterative scheme \sep Fast convergence \sep Asymptotic preserving \sep High-order finite element method
\end{keywords}

\maketitle

\section{Introduction}
\label{sec: introduction}
In this work, we develop a synthetic iterative scheme to solve the stationary, linearized, non-gray Callaway model for multiscale phonon heat transport~\cite{callaway1959model}. It is a dual-relaxation-time approximation to the phonon Boltzmann transport equation (PBTE), separating the distinct roles of resistive and normal scattering processes: resistive scattering drives the phonon distribution function toward a Bose-Einstein equilibrium, whereas normal scattering relaxes it toward the drifted Bose-Einstein equilibrium. In addition, non-gray model retains the branch- and frequency-dependent phonon spectrum, taking into account contributions from different phonon modes. The non-gray Callaway equation provides a comprehensive kinetic framework for predicting phonon transport across diffusive, ballistic, hydrodynamic, and Ziman regimes~\cite{chen2021nonfourier}, which is imperative for thermal management of microelectronic devices or understanding physics in new materials when the classical Fourier's law is no longer valid~\cite{bao2018review,guo2015hydrodynamics}.

For a thermodynamic system with a small temperature deviation from a global reference temperature $T_0$, the linearized Callaway equation reads
\begin{equation}\label{eq: BTE_energy}
    \frac{\partial e}{\partial t} + \bm{v}\cdot\nabla_{\bm{x}} e = \frac{e_{\mathrm{R}}^{\mathrm{eq}}(T_{\mathrm{R}}^{\mathrm{loc}}) - e}{\tau_{\mathrm{R}}} + \frac{e_{\mathrm{N}}^{\mathrm{eq}}(T_{\mathrm{N}}^{\mathrm{loc}}, \bm{u}) - e}{\tau_{\mathrm{N}}},
\end{equation}
where, $e=e(\bm{x},t,p,\omega,\bm{s})$ denotes the phonon deviated energy distribution function for a mode with polarization branch $p$ and angular frequency $\omega$ in a propagation direction $\bm{s}$ at spatial position $\bm{x}=(x,y)$ and time $t$; $\bm{v}=\nabla_{\bm{k}}\omega$ is the group velocity associated with the dispersion relation $\omega(\bm{k})$ and the wave vector $\bm{k}$. Due to isotropy, the dispersion relation depends solely on the magnitude of the wave vector, that is, $\omega=\omega(p,|\bm{k}|)$. Consequently, $\bm{v}\parallel\bm{k}$, and we write $\bm{v}=|\bm{v}|\bm{s}$ and $\bm{k}=|\bm{k}|\bm{s}$; the direction $\bm{s}\in S:=\{\bm{w}\in\mathbb{R}^d:\|\bm{w}\|=1\}$ is an ordinary unit vector with $d$ being the number of dimensions. $\tau_{\mathrm{R}}(p,\omega)$ and $\tau_{\mathrm{N}}(p,\omega)$ are the mode-dependent relaxation times responsible for the resistive and normal scattering processes, respectively. The equilibria $e_{\mathrm{R}}^{\mathrm{eq}}$ and $e_{\mathrm{N}}^{\mathrm{eq}}$ are written in
\begin{subequations}\label{eq: e_R_N_lin}
\begin{align}
    e_{\mathrm{R}}^{\mathrm{eq}}(T_{\mathrm{R}}^{\mathrm{loc}}) &\simeq \frac{C_{p,\omega}}{\mathrm{B}}(T_{\mathrm{R}}^{\mathrm{loc}} - T_0), \label{eq: e_R_lin}\\
    e_{\mathrm{N}}^{\mathrm{eq}}(T_{\mathrm{N}}^{\mathrm{loc}}, \bm{u}) &\simeq \frac{C_{p,\omega}}{\mathrm{B}}\left[(T_{\mathrm{N}}^{\mathrm{loc}} - T_0) + T_0\frac{\bm{k}\cdot\bm{u}}{\omega}\right], \label{eq: e_N_lin}
\end{align}
\end{subequations}
where the mode-specific heat capacity $C_{p,\omega}$ is defined as
\begin{equation}\label{eq: Cpomega}
    C_{p,\omega} = \frac{\hbar^2\omega^2 D \exp\{\hbar\omega/(k_B T_0)\}}{\bigl(\exp\{\hbar\omega/(k_B T_0)\}-1\bigr)^2 k_B T_0^2};
\end{equation}
$B=2\pi$ ($4\pi$) and the density state $D=|\bm{k}|/2\pi|\bm{v}|$ ($|\bm{k}|^2/2\pi^2|\bm{v}|$) for two-dimensional (three-dimensional) isotropic materials; $\hbar$ and $k_B$ are the reduced Planck and the Boltzmann constants, respectively. The equilibria involve three quantities, namely, pseudo-temperatures $T_{\mathrm{R}}^{\mathrm{loc}}$ and $T_{\mathrm{N}}^{\mathrm{loc}}$ that enforce energy conservation for the resistive and normal processes, respectively; drift velocity $\bm{u}$ determined by conservation of phonon quasi-momentum under normal scattering. They are evaluated as
\begin{subequations}\label{eq: T_loc_drift_vel}
\begin{align}
    T_{\mathrm{R}}^{\mathrm{loc}} &= T_0 + \frac{1}{C_{\tau_{\mathrm{R}}}}\left\langle \frac{e}{\tau_{\mathrm{R}}}\right\rangle_{c}, \qquad C_{\tau_{\mathrm{R}}} = \left\langle \frac{C_{p,\omega}}{\tau_{\mathrm{R}}} \right\rangle_{p}, \label{eq: T_loc_R}\\
    T_{\mathrm{N}}^{\mathrm{loc}} &= T_0 + \frac{1}{C_{\tau_{\mathrm{N}}}}\left\langle \frac{e}{\tau_{\mathrm{N}}}\right\rangle_{c}, \qquad C_{\tau_{\mathrm{N}}} = \left\langle \frac{C_{p,\omega}}{\tau_{\mathrm{N}}} \right\rangle_{p}, \label{eq: T_loc_N}\\
    \bm{u} &= \frac{d}{T_0 C_{\tau_{\mathrm{N}}}^{1}} \left\langle \frac{\bm{k}}{\omega} \frac{e}{\tau_{\mathrm{N}}}\right\rangle_{c}, \qquad C_{\tau_{\mathrm{N}}}^{1} = \left\langle \frac{|\bm{k}|^2}{\omega^2} \frac{C_{p,\omega}}{\tau_{\mathrm{N}}} \right\rangle_{p}, \label{eq: drift_vel}
\end{align}
\end{subequations}
where $\langle\cdot\rangle_{c}:=\sum_p\iint\cdot\,\mathrm{d}\Omega\mathrm{d}\omega$ and $\langle\cdot\rangle_{p}:=\sum_p\int\cdot\,\mathrm{d}\omega$ with $\mathrm{d}\Omega$ the differential plane (solid) angle. Other macroscopic quantities of interest, including deviated energy density $E$, heat capacity $C_V$, equivalent equilibrium temperature $T$, defined as~\cite{chen2005nanoscale} 
\begin{equation}\label{eq: energy_dens_T}
    E = C_V \bigl(T - T_0\bigr) = \left\langle e \right\rangle_{c}, \qquad T=T_0+\frac{1}{C_V}\left\langle e \right\rangle_{c},\qquad C_V = \left\langle C_{p,\omega} \right\rangle_{p},
\end{equation}
as well as heat flux $\bm{q}$, momentum-flux tensor $\bm{\Pi}$, momentum density $\bm{p}$, and resistive momentum source $\bm{p}_{\mathrm{R}}$
\begin{equation}\label{eq: heat_momentum_flux}
    \bm{q} = \left\langle \bm{v} \, e \right\rangle_{c},\quad \bm{\Pi}(\bm{x},t) = \left\langle \frac{\bm{v}\otimes\bm{k}}{\omega} \, e \right\rangle_{c},\quad \bm{p}(\bm{x},t) = \left\langle \frac{\bm{k}}{\omega} \, e \right\rangle_{c},\quad \bm{p}_{\mathrm{R}} = \left\langle \frac{\bm{k}}{\omega\tau_{\mathrm{R}}} \, e \right\rangle_{c}.
\end{equation}
Under isotropic dispersion, $v_i k_j=v_j k_i$, so $\bm{\Pi}$ is symmetric.


Various numerical methods have been developed for PBTE, nevertheless, several challenges remain \cite{bao2018review,Mazumder2022}. Although energy-based variance-reduced Monte Carlo methods efficiently suppress statistical noise for small deviations from equilibrium and provide considerable flexibility in high-dimensional phase space, sufficiently accurate local fields still require adequate sampling \cite{peraud2011efficient}. Deterministic approaches, combining the discrete ordinate method (DOM) and finite-volume/difference/element or spherical-harmonic discretizations \cite{murthy2002computation,mittal2011hybrid,shen2025accurate,luo2025dg}, provide noise-free solutions to boundary-value problems, but their computational costs grow with the product of spatial, angular and spectral resolutions. Another finite-volume framework named unified gas kinetic methods is created in particular for multiscale simulation \cite{guo2016dugks,luo2017dugks,liu2025ugkwp,liu2025ugkwpiugkp}. The methods overcome the stiffness in the diffusion limit by constructing a multiscale interface flux from time-space (transport-scattering) coupled evolution. More recently, physics-informed neural networks were applied to parameterized PBTE problems with JAX-BTE providing a GPU-accelerated framework for forward and inverse calculations \cite{li2021pinn,zhou2023pinn,shang2025jaxbte}. However, precision and robustness are sensitive to optimisation, residual balancing, and collocation sampling \cite{wang2022pinn}, while JAX-BTE retains the memory costs of deterministic discretization. 

A kinetic method usually uses iterative or time marching schemes for stationary solutions, among which the source iteration~\cite{adams2002fast} forms the conventional iterative scheme (CIS). The scheme treats transport and collision-loss terms implicitly, while the collision-gain term is evaluated from the solution obtained in the preceding iteration. The treatment avoids a monolithic inversion of all phase-space unknowns, but the macroscopic fields are updated through successive kinetic sweeps. In the ballistic regime, a kinetic sweep can substantially reduce iteration errors and CIS converges rapidly. However, as the material size becomes large over the relaxation lengths of many phonons, errors that in the material scale are reduced slowly by the local iteration; consequently, the convergence of CIS is significantly dragged down. Works on non-gray PBTE solvers have reported poor convergence when the mode-resolved Knudsen numbers become small \cite{loy2015coupled}. This motivates the so-called synthetic acceleration \cite{adams2002fast}, which follows the principle: macroscopic equations are solved together with the kinetic equation so that the slowly deteriorating error is corrected globally \cite{wu2017fast,su2020can,su2020fast}. Nevertheless, the form of the macroscopic synthetic equation depends on the collision invariants and the spectral representation of the kinetic model.


\subsection{Related works}
The works most relevant to our development on efficiently solving non-gray and/or dual-relaxation problems can be organized into three lines.

\emph{Non-gray, single-relaxation synthetic schemes.} Zhang et al.~derived a diffusion-type synthetic equation from energy conservation to accelerate the iterative solution, in particular in/near the diffusion transport regime, of stationary non-gray PBTE under a single-relaxation-time model \cite{zhang2021synthetic}. Subsequent acceleration strategies and implementations, including GiftBTE \cite{hu2024giftbte} and extension to large temperature variation \cite{zhang2025hotspot}, apply macroscopic corrections based on temperature or energy for spectrum-dependent phonon properties \cite{zhang2023acceleration,chen2026nanobte}. 

\emph{Gray, dual-relaxation synthetic scheme.} Liu et al.~constructed synthetic equations for the gray Callaway equation and demonstrated rapid convergence in both the diffusive and hydrodynamic limits \cite{liu2022fast}. This work establishes the synthetic macroscopic structure required when normal and resistive scattering processes are distinguished, but the phonon properties and relaxation times are represented by gray, mode-independent quantities rather than a branch- and frequency-resolved spectrum.
    
\emph{Direct non-gray, dual-relaxation solvers.} Guo and Wang directly solved the non-gray Callaway equation for graphene using DOM and finite-difference spatial discretization. More recently, discontinuous Galerkin finite element methods extended such calculations to irregular geometries partitioned by unstructured meshes \cite{guo2017heat,luo2025dg,luo2025porous}. These methods retain both the spectral dependence and the separate normal and resistive collision channels but are generally based on CIS.

Each line combines two of the three major components consisting of the current work: spectrum-resolved transport, dual-relaxation physics, and synthetic acceleration, but not all three. A general synthetic iterative scheme (GSIS) has not yet been formulated for the non-gray Callaway PBTE with dual-relaxation approximations. Such a development cannot be obtained by simply replacing the gray coefficients by spectral averages in the synthetic scheme for the gray Callaway model: the collision source contains two local pseudo-temperatures and a drift velocity, while the macroscopic equations must consistently retain the branch- and frequency-dependent dispersion and both relaxation channels. The Guyer-Krumhansl equation \cite{guyer1966solution,guyer1966thermal} and the extension to take into account quadratic dispersion \cite{Shang2020HydroPhonon}, as well as the Grad's moment equations~\cite{Fryer2014moment} are macroscopic equations derived from the Callaway model. They are obtained on the basis of gray assumption and truncation of the phonon distribution function, thus, cannot be used for synthetic approximation in non-gray, multiscale simulations. Constructing a closed synthetic macroscopic system is therefore the central step in extending GSIS to non-gray dual-relaxation problems. 

\subsection{Our contributions} We develop a GSIS for stationary, linearized non-gray Callaway PBTE. Major contributions include the following.

\emph{Derive synthetic macroscopic equations and iterative scheme:} a Chapman--Enskog analysis~\cite{chapman1990mathematical} for energy and quasi-momentum balance laws is used to derive first-order constitutive relations for the heat flux, momentum-flux tensor, etc. Higher-order, non-equilibrium terms evaluated from the kinetic solution, yielding a closed system for the pseudo-temperatures and drift velocity covering the whole transport regimes, whose solution updates the collision source in the kinetic iteration.

\emph{Achieve high-order discretization in the spatial space:} the phonon spectrum and direction resolved kinetic equation is discretized by an upwind nodal DG method, while the synthetic equations are solved by a hybridizable discontinuous Galerkin (HDG) method, with a consistent treatment of physical boundary data. The finite element system provides high-order discretization and flexible treatment of complex geometries. 
    
\emph{Perform rigorous stability and asymptotic analyses:} a branch- and frequency-resolved Fourier analysis is used to characterize the convergence of CIS and GSIS, while an asymptotic analysis based on the Chapman-Enskog expansion is applied to show the asymptotic behavior of the proposed numerical method in the diffusion and hydrodynamic transport regimes.

The paper is organized as follows: \Cref{sec: gsis_formu_alg} introduces the formulations and the complete algorithm of GSIS. Corresponding Fourier convergence analysis and the asymptotic preserving analysis are given in \Cref{sec: fourier_analysis} and \Cref{sec: ap_analysis}, respectively. In \Cref{sec: dg_hdg}, the spatial discretization schemes by DG and HDG are demonstrated. \Cref{sec: numerical_validation} describes the numerical validation and test examples over a particular graphene phonon model. Conclusions and outlook are summarized in \Cref{sec: conclusion_outlook}.

\section{GSIS: formulation and algorithm}
\label{sec: gsis_formu_alg}

The synthetic equations start from the exact energy conservation and the quasi-momentum balance law, read
\begin{subequations}\label{eq: exact_balance}
    \begin{align}
        \frac{\partial E}{\partial t} + \nabla_{\bm{x}}\cdot\bm{q} &= 0, \label{eq: energy_consv_law}\\
        \frac{\partial\bm{p}}{\partial t} + \nabla_{\bm{x}}\cdot\bm{\Pi} &= -\bm{p}_{\mathrm{R}}, \label{eq: momentum_consv_law}
    \end{align}
\end{subequations}
which are obtained by taking $\langle\cdot\rangle_{c}$ to Eq. \eqref{eq: BTE_energy} and the resulting equation to multiply Eq. \eqref{eq: BTE_energy} by $\bm{k}/\omega$. They can be understood as the conservation laws associated with collision invariants $T_{\mathrm{N}}^{\mathrm{loc}}$ and $\bm{u}$ in normal scattering. The equations are not closed. Constitutive relations are required for the fluxes $\bm{q}$ and $\bm{\Pi}$, as well as the source term $\bm{p}_{\mathrm{R}}$. We construct them as the first-order Chapman-Enskog approximations in terms of $T_{\mathrm{N}}^{\mathrm{loc}}$ and $\bm{u}$ plus fully high-order, non-equilibrium terms. In addition, a formulation is needed to retrieve the field of $T_{\mathrm{R}}^{\mathrm{loc}}$ because it also appears explicitly in the equilibria of the Callaway model. 

\subsection{Chapman-Enskog analysis} 
\label{sec: gsis_ce_analysis}
Considering the hydrodynamic limit when normal scattering dominates whereas resistive scattering is restricted, a small parameter $\epsilon$ can be used to scale the kinetic equation as
\begin{equation}\label{eq: CE_BTE_hydro}
    \frac{\partial e}{\partial t} + \bm{v}\cdot\nabla_{\bm{x}} e = \epsilon\,\frac{e_{\mathrm{R}}^{\mathrm{eq}} - e}{\widehat{\tau}_{\mathrm{R}}} + \frac{e_{\mathrm{N}}^{\mathrm{eq}} - e}{\epsilon\widehat{\tau}_{\mathrm{N}}},
\end{equation}
where $\widehat{\tau}_{\mathrm{R}} = \epsilon\tau_{\mathrm{R}}$ and $\widehat{\tau}_{\mathrm{N}} = \tau_{\mathrm{N}}/\epsilon$. Then, we can expand the distribution function and the time derivative over $\epsilon$ as
\begin{equation}\label{eq: CE_expansion}
    e = e^{(0)} + \epsilon\, e^{(1)} + \epsilon^2\, e^{(2)} + \cdots, \qquad\partial_t = \partial_{t_0} + \epsilon \partial_{t_1} + \epsilon^2 \partial_{t_2} + \cdots.
\end{equation}
We want to obtain the up to first-order approximations $\bm{M}_{\mathrm{ap}}$ to the macroscopic quantities, say $\bm{M}\simeq\bm{M}_{\mathrm{ap}}:=\bm{M}^{(0)}+\epsilon\bm{M}^{(1)}$, where $\bm{M}:=\{T_{\mathrm{R}}^{\mathrm{loc}},\bm{q},\bm{\Pi},\bm{p}_{\mathrm{R}}\}$, and the coefficients $\bm{M}^{(i)}$ are evaluated from $e^{(i)}$ according to Eqs. \eqref{eq: energy_dens_T} and \eqref{eq: heat_momentum_flux}. The pseudo-temperature $T_{\mathrm{N}}^{\mathrm{loc}}$ and the drifted velocity $\bm{u}$ are collision invariants in normal scattering and are not expanded~\cite{chapman1990mathematical}. Multiplying \eqref{eq: CE_BTE_hydro} by $\epsilon$, substituting expansion~\eqref{eq: CE_expansion}, and collecting equal powers of $\epsilon$, we have
\begin{subequations}
    \begin{align}
    \epsilon^{0}&: \quad e_{\mathrm{N}}^{\mathrm{eq}} - e^{(0)} = 0, \label{eq: hydro_eps_m1}\\[4pt]
    \epsilon^{-1}&: \quad \partial_{t_0} e^{(0)} + \bm{v}\cdot\nabla_{\bm{x}}e^{(0)} = -\frac{e^{(1)}}{\widehat{\tau}_{\mathrm{N}}}.\label{eq: hydro_eps_0}
\end{align}
\end{subequations}

\paragraph{Zeroth-order solution}
Equation~\eqref{eq: hydro_eps_m1} gives the local drifted equilibrium
\begin{equation}\label{eq: CE_0th}
    e^{(0)} = e_{\mathrm{N}}^{\mathrm{eq}} = \frac{C_{p,\omega}}{\mathrm{B}}\left[(T_{\mathrm{N}}^{\mathrm{loc}} - T_0) + T_0\frac{\bm{k}\cdot\bm{u}}{\omega}\right].
\end{equation}
Its macroscopic moments satisfy the following.
\begin{equation}\label{eq: CE_0th_macro}
    \quad (T_{\mathrm{R}}^{\mathrm{loc}})^{(0)} = T_{\mathrm{N}}^{\mathrm{loc}}, \quad \bm{q}^{(0)} = \frac{T_0}{d}\,C_{vk}\,\bm{u},
    \quad  
    \bm{\Pi}^{(0)} = \frac{C_{vk}}{d}(T_{\mathrm{N}}^{\mathrm{loc}} - T_0)\,\bm{I}, 
    \quad
    \bm{p}_{\mathrm{R}}^{(0)} = \frac{T_0}{d}\,C_{\tau_{\mathrm{R}}}^{1}\,\bm{u},
\end{equation}
where $\bm{I}$ is the unit matrix. It is easy to see $\left\langle e^{(0)}/{\tau_{\mathrm{N}}}\right\rangle_{c}/C_{\tau_{\mathrm{N}}}=T_{\mathrm{N}}^{\mathrm{loc}} - T_0 $ and $d\left\langle \bm{k}e/{\omega\tau_{\mathrm{N}}}\right\rangle_{c}/T_0 C_{\tau_{\mathrm{N}}}^{1}=\bm{u}$, therefore, we can impose the following compatible condition for the energy and quasi-momentum conservation of normal scattering processes.
\begin{equation}\label{eq: hydro_compat}
    \left\langle \frac{e^{(j)}}{\tau_{\mathrm{N}}} \right\rangle_c = 0, \qquad
    \left\langle \frac{\bm{k}}{\omega}\frac{e^{(j)}}{\tau_{\mathrm{N}}} \right\rangle_c = \bm{0}, \qquad j=1,2,\ldots.
\end{equation}

\paragraph{First-order solution}
From \eqref{eq: hydro_eps_0} and \eqref{eq: CE_0th},
\begin{equation}\label{eq: e1_raw}
    \frac{e^{(1)}}{\widehat{\tau}_{\mathrm{N}}} = -\frac{C_{p,\omega}}{\mathrm{B}}\left[\partial_{t_0}(T_{\mathrm{N}}^{\mathrm{loc}} - T_0) + \frac{T_0}{\omega}\bm{k}\cdot\partial_{t_0}\bm{u} + \bm{v}\cdot\nabla_{\bm{x}}(T_{\mathrm{N}}^{\mathrm{loc}} - T_0) + \frac{T_0}{\omega}\bigl(\bm{v}\otimes\bm{k}:\nabla_{\bm{x}}\bm{u}\bigr)\right],
\end{equation}
where $\bm{A}:\bm{B}=\sum_{i,j}A_{ij}B_{ij}$ denotes the Frobenius inner product of matrices. The terms of time derivatives are eliminated by the corresponding energy and quasi-momentum balance laws of Eq. \eqref{eq: hydro_eps_0} and using the condition \eqref{eq: hydro_compat}, resulting in
\begin{equation}\label{eq: dt0_explicit}
    \partial_{t_0}(T_{\mathrm{N}}^{\mathrm{loc}} - T_0) = -\frac{T_0 C_{vk}}{d C_V}\,\nabla_{\bm{x}}\cdot\bm{u}, \quad
    \partial_{t_0}\bm{u} = -\frac{C_{vk}}{T_0 C_k}\,\nabla_{\bm{x}}(T_{\mathrm{N}}^{\mathrm{loc}} - T_0). 
\end{equation}
Then, we have
\begin{equation}\label{eq: e1_solution}
    e^{(1)} = -\widehat{\tau}_{\mathrm{N}}\frac{C_{p,\omega}}{\mathrm{B}}\left[\frac{T_0}{\omega}\bigl(\bm{v}\otimes\bm{k}:\nabla_{\bm{x}}\bm{u}\bigr) - \frac{T_0 C_{vk}}{d C_V}\nabla_{\bm{x}}\cdot\bm{u} + \left(\bm{v} - \frac{C_{vk}}{C_k}\frac{\bm{k}}{\omega}\right)\cdot\nabla_{\bm{x}}(T_{\mathrm{N}}^{\mathrm{loc}} - T_0)\right].
\end{equation}
Using \eqref{eq: e1_solution}, we can calculate $\bm{M}^{(1)}$, and eventually we obtain the up to first-order approximations, read
\begin{subequations}\label{eq: CE_CR}
    \begin{align}
        \bm{q}_{\mathrm{ap}}=& \frac{T_0 C_{vk}}{d}\,\bm{u} - \kappa_{\mathrm{N}}\,\nabla_{\bm{x}}(T_{\mathrm{N}}^{\mathrm{loc}} - T_0), \label{eq: CE_q}\\
        \bm{\Pi}_{\mathrm{ap}}=&\frac{C_{vk}}{d}\,(T_{\mathrm{N}}^{\mathrm{loc}} - T_0)\,\bm{I} - \mu_{\mathrm{N}}^{1}\left[\nabla_{\bm{x}}\bm{u} + (\nabla_{\bm{x}}\bm{u})^{\top} + (\nabla_{\bm{x}}\cdot\bm{u})\bm{I}\right]+ \mu_{\mathrm{N}}^{2}\,(\nabla_{\bm{x}}\cdot\bm{u})\bm{I}, \label{eq: CE_Pi}\\
        \bm{p}_{\mathrm{R,ap}}=&\frac{T_0 C_{\tau_{\mathrm{R}}}^{1}}{d}\,\bm{u} - \eta\,\nabla_{\bm{x}}(T_{\mathrm{N}}^{\mathrm{loc}} - T_0), \label{eq: CE_pR}\\
        T_{\mathrm{R,ap}}^{\mathrm{loc}}=& T_{\mathrm{N}}^{\mathrm{loc}} + \gamma\,\nabla_{\bm{x}}\cdot\bm{u}. \label{eq: CE_TR}
    \end{align}
\end{subequations}
The transport coefficients in Eqs.~\eqref{eq: CE_0th_macro} and~\eqref{eq: CE_CR} are given in Appendix \ref{sec: A1}.

\subsection{Full constitutive relations}
Since we aim at multiscale simulations, requiring coverage of the whole transport regime, the high-order terms $\mathrm{HoT}_{\bm{M}}$ taking into account non-equilibrium effects should be included, that is, 
\begin{equation}
    \bm{M}=\bm{M}_{\mathrm{ap}}+\mathrm{HoT}_{\bm{M}}.
\end{equation}
The high-order terms are modeled as the residuals of the first-order constitutive relations and will be detailed in the next section.

\subsection{GSIS algorithm}
\label{sec: GSIS_scheme}

One GSIS iteration maps $\{e^n,\bm{M}^n\}$ to $\{e^{n+1},\bm{M}^{n+1}\}$ through a kinetic solver and a macroscopic synthetic solver. Without confusion, in this section, the collection of macroscopic quantities is $\bm{M}:=\{T_{\mathrm{R}}^{\mathrm{loc}},T_{\mathrm{N}}^{\mathrm{loc}},T,\bm{u},\bm{q}\}$. The iteration is split into two steps. First, an intermediate solution $e^{n+1/2}$ is evaluated from
\begin{equation}\label{eq: GSIS_kinetic_half_step}
    \bm v\cdot\nabla_{\bm x}e^{n+1/2}=\frac{e_{\mathrm R}^{\mathrm{eq}}(T_{\mathrm R}^{\mathrm{loc},n})-e^{n+1/2}}{\tau_{\mathrm R}}+\frac{e_{\mathrm N}^{\mathrm{eq}}(T_{\mathrm N}^{\mathrm{loc},n},\bm u^n)-e^{n+1/2}}{\tau_{\mathrm N}},
\end{equation}
Using which, the intermediate macroscopic quantities $\bm{M}^{n+1/2}$ are obtained by taking integrals of $e^{n+1/2}$, see Eqs.~\eqref{eq: T_loc_drift_vel}--\eqref{eq: heat_momentum_flux}. The high-order terms are also calculated by this intermediate kinetic solution, modeling as the residuals of the first-order constitutive relations \eqref{eq: CE_CR} as follows
\begin{subequations}\label{eq: HoT_defs}
\begin{align}
    \mathrm{HoT}_{\bm{q}}^{n+1/2}
    &= \bm{q}^{n+1/2} - \frac{T_0 C_{vk}}{d}\,\bm{u}^{n+1/2} + \kappa_{\mathrm{N}}\nabla_{\bm{x}}(T_{\mathrm{N}}^{\mathrm{loc},n+1/2} - T_0), \label{eq: HoT_q}\\
    \mathrm{HoT}_{\bm{\Pi}}^{n+1/2}
    &= \bm{\Pi}^{n+1/2} - \frac{C_{vk}}{d}(T_{\mathrm{N}}^{\mathrm{loc},n+1/2}-T_0)\bm{I} \nonumber\\
    &\quad + \mu_{\mathrm{N}}^{1}\left[\nabla_{\bm{x}}\bm{u}^{n+1/2}+(\nabla_{\bm{x}}\bm{u}^{n+1/2})^{\top}+(\nabla_{\bm{x}}\cdot\bm{u}^{n+1/2})\bm{I}\right] - \mu_{\mathrm{N}}^{2}(\nabla_{\bm{x}}\cdot\bm{u}^{n+1/2})\bm{I},  \label{eq: HoT_Pi}\\
    \mathrm{HoT}_{\bm{p}_\mathrm{R}}^{n+1/2}
    &= \bm{p}_{\mathrm{R}}^{n+1/2} - \frac{T_0 C_{\tau_{\mathrm{R}}}^{1}}{d}\,\bm{u}^{n+1/2} + \eta\nabla_{\bm{x}}(T_{\mathrm{N}}^{\mathrm{loc},n+1/2}-T_0), \label{eq: HoT_p}\\
    \mathrm{HoT}_{T_{\mathrm{R}}^{\mathrm{loc}}}^{n+1/2}
    &= T_{\mathrm{R}}^{\mathrm{loc},n+1/2} - T_{\mathrm{N}}^{\mathrm{loc},n+1/2} - \gamma\nabla_{\bm{x}}\cdot\bm{u}^{n+1/2}. \label{eq: HoT_T}
\end{align}
\end{subequations}
Then, the synthetic approximations for the conservative quantities $T_{\mathrm{N}}^{\mathrm{loc},n+1}$ and $\bm{u}^{n+1}$ are solved from the following macroscopic equations that are obtained by inserting the fully constitutive laws for $\bm{q}$, $\bm{\Pi}$, and $\bm{p}_{\mathrm{R}}$ into the exact balance laws~\eqref{eq: exact_balance}, 
\begin{subequations}\label{eq: synth_energy_momentum}
    \begin{align}
    &\frac{T_0 C_{vk}}{d}\,\nabla_{\bm{x}}\cdot\bm{u}^{n+1} - \kappa_{\mathrm{N}}\,\Delta_{\bm{x}}(T_{\mathrm{N}}^{\mathrm{loc},n+1}-T_0) = -\nabla_{\bm{x}}\cdot\mathrm{HoT}_{\bm{q}}^{n+1/2}, \label{eq: synth_energy_momentum_a}\\
    &\left(\frac{C_{vk}}{d}-\eta\right)\nabla_{\bm{x}}(T_{\mathrm{N}}^{\mathrm{loc},n+1}-T_0)- \mu_{\mathrm{N}}^{1}\left[\Delta_{\bm{x}}\bm{u}^{n+1}+2\nabla_{\bm{x}}(\nabla_{\bm{x}}\cdot\bm{u}^{n+1})\right] \nonumber \\
    &\qquad\qquad + \mu_{\mathrm{N}}^{2}\nabla_{\bm{x}}(\nabla_{\bm{x}}\cdot\bm{u}^{n+1})
    + \frac{T_0 C_{\tau_{\mathrm{R}}}^{1}}{d}\bm{u}^{n+1}=-\nabla_{\bm{x}}\cdot\mathrm{HoT}_{\bm{\Pi}}^{n+1/2} - \mathrm{HoT}_{\bm{p}_{\mathrm{R}}}^{n+1/2}, \label{eq: synth_energy_momentum_b}
    \end{align}
\end{subequations}
while the other synthetic fields are retrieved from $T_{\mathrm{N}}^{\mathrm{loc},n+1}$ and $\bm{u}^{n+1}$ by
\begin{subequations}\label{eq: update_T_q}
\begin{align}
    T_{\mathrm{R}}^{\mathrm{loc},n+1} &= T_{\mathrm{N}}^{\mathrm{loc},n+1} + \gamma\nabla_{\bm{x}}\cdot\bm{u}^{n+1} + \mathrm{HoT}_{T_{\mathrm{R}}^{\mathrm{loc}}}^{n+1/2}, \label{eq: update_T_R}\\
    T^{n+1} &= T_{\mathrm{N}}^{\mathrm{loc},n+1} + \frac{1}{C_V}\left\langle e^{n+1/2}-e_{\mathrm N}^{\mathrm{eq}}\left(T_{\mathrm{N}}^{\mathrm{loc},n+1/2},\bm u^{n+1/2}\right)\right\rangle_{c}, \label{eq: update_T_mac}\\
    \bm{q}^{n+1} &= \frac{T_0 C_{vk}}{d}\,\bm{u}^{n+1} - \kappa_{\mathrm{N}}\,\nabla_{\bm{x}}(T_{\mathrm{N}}^{\mathrm{loc},n+1} - T_0) + \mathrm{HoT}_{\bm{q}}^{n+1/2}. \label{eq: update_q_mac}
\end{align}
\end{subequations}
Given a user-defined tolerance $\mathrm{tol}$, the criterion for iteration termination is set as
\begin{equation}\label{eq: stop_criterion}
   \varepsilon_T:=\frac{\|T^{n+1}-T^{n}\|_2}{\|T^{n+1}\|_2}<\mathrm{tol},
   \qquad
   \varepsilon_q:=\frac{\|\bm q^{n+1}-\bm q^{n}\|_2}{\|\bm q^{n+1}\|_2}<\mathrm{tol},
\end{equation}
where $\|\cdot\|_2$ is the standard $L_2$ norm. The algorithm \ref{alg: gsis} demonstrates the complete GSIS iteration cycles. CIS iteration can be simply understood by skipping the synthetic macroscopic solutions (line 5--7) and setting $\bm{M}^{n+1}=\bm{M}^{n+1/2}$.


\begin{algorithm}[htbp]
\caption{General synthetic iterative scheme for the stationary non-gray Callaway equation}
\label{alg: gsis}
\begin{algorithmic}[1]
\Require $e^0$, $\bm M^0$, $\mathrm{tol}$
\For{$n=0,1,2,\ldots$}
    \State Construct $e_{\mathrm R}^{\mathrm{eq},n}$ and $e_{\mathrm N}^{\mathrm{eq},n}$ from $\bm M^n$
    \State Solve Eq. \eqref{eq: GSIS_kinetic_half_step} for $e^{n+1/2}$
    \State Evaluate $\bm M^{n+1/2}$ by taking moments of $e^{n+1/2}$
    \State Evaluate the high-order terms by Eq. \eqref{eq: HoT_defs}
    \State Solve Eq. \eqref{eq: synth_energy_momentum} for $T_{\mathrm N}^{\mathrm{loc},n+1}$ and $\bm u^{n+1}$
    \State Reconstruct $\bm{q}^{n+1}$, $T_{\mathrm{R}}^{\mathrm{loc},n+1}$ and $T^{n+1}$ using Eq. \eqref{eq: update_T_q}
    \State Set $e^{n+1}=e^{n+1/2}$
    \State Compute $\varepsilon_T$ and $\varepsilon_q$ from Eq. \eqref{eq: stop_criterion}
    \If{$\varepsilon_T<\mathrm{tol}$ and $\varepsilon_q<\mathrm{tol}$}
        \Return $e^{n+1}$ and $\bm M^{n+1}$
    \EndIf
\EndFor
\end{algorithmic}
\end{algorithm}

\section{Fourier convergence analysis}
\label{sec: fourier_analysis}

In this section, we use Fourier's analysis to show the convergence property of CIS and GSIS for stationary solutions of Eq. \eqref{eq: BTE_energy}. Continuous-space analysis with periodic boundary condition follows the error associated with a single spatial Fourier mode and identifies the corresponding amplification matrix, whose spectral radius gives the asymptotic contraction factor of that mode \cite{adams2002fast,liu2022fast}. We specialize in $d=2$, write
\[
    \bm x=(x,y),
    \qquad
    \bm s=(\cos\theta,\sin\theta):=(\mu,\zeta),
\]
where $\theta$ is the polar angle that forms the directional vector. We define the combined relaxation time and the phonon-spectrum-dependent relaxation length as
\begin{equation}\label{eq: tauC_def}
    \tau_{\mathrm C}^{-1}(p,\omega)
    :=\tau_{\mathrm R}^{-1}(p,\omega)+\tau_{\mathrm N}^{-1}(p,\omega),
    \qquad
    \ell_{p,\omega}:=v_{p,\omega}\tau_{\mathrm C}(p,\omega).
\end{equation}

Let $a^\star$ denote the fixed point of any field $a$ and define its $n$-th iteration error with the Fourier ansatz as 
\[
\tilde a^{\,n} := a^n-a^\star = \sum_{z\in\mathbb{Z}} \hat{a}_{z}^n \exp(\mathrm i\bm\xi_z\cdot\bm x), \qquad \bm\xi_z\in\mathbb R^2,
\]
where $\hat{a}_{z}^n$ is the amplitude of $z$-th error mode. For any nonzero mode, $\bm\xi_z$ is the vector of wavenumbers defined by
\begin{equation}\label{eq: error_wavenumber}
    \bm \xi_z = 2\pi\left(\frac{z}{L}, \frac{z}{W}\right)^{\top},
\end{equation}
where $L$ and $W$ are longitudinal and transverse sizes of a material, respectively. Rotational isotropy makes the Fourier symbol a function of $\xi := |\bm \xi|$ only. Without loss of generality, we choose the wave vector aligned with the $x$-axis, that is, $\bm \xi = \xi (1, 0)^{\top}$, $\xi \geq 0$. 
Due to linearity of the problem, one can treat one Fourier mode at a time, that is, perform a single-mode convergence diagnostic. We choose $z = 1$ and then sweeping over $z$ could be converted to sweeping over $L$. Therefore, we omit the subscripts $z$ for simplicity in the following analysis. In particular, 
\begin{equation}\label{eq: ansatz}
    \tilde a^{n}
    =\hat a^{\,n+1/2}
     \exp(\mathrm i\bm\xi\cdot\bm x),
    \qquad
    \bm\xi = \left(\frac{2\pi}{L}, 0\right)^{\top}\in\mathbb R^2.
\end{equation}
We are interested in the following four amplitudes of macroscopic quantities
\begin{equation}\label{eq: M_iter_err_amp}
    \hat{\bm M}^n := \left(\hat T_{\mathrm R}^{\,n},\,\ \hat T_{\mathrm N}^{\,n},\,\ \hat u_x^{\,n},\,\ \hat u_y^{\,n}\right)^{\!\top}\in\mathbb C^4.
\end{equation} 
and obtain the error propagation map $\hat{\bm M}^n\rightarrow\hat{\bm M}^{n+1}$ in both CIS and GSIS. With $\bm\xi\parallel (1, 0)^{\top}$, the longitudinal variables $(\hat T_{\mathrm R},\hat T_{\mathrm N},\hat u_x)$ decouple from the transverse variable $\hat u_y$. The amplification matrices consequently have a block structure of $3\times3\oplus1$.

\subsection{Contraction matrices}\label{sec: fourier_cis_gsis}

\paragraph{Contraction matrix of CIS}
At the $n+1$-th iteration of CIS, the single mode error of the distribution function is 
\begin{equation}\label{eq: e_ansatz}
    \tilde e^{\,n+1}(x,p,\omega,\theta)
    =\hat e^{\,n+1}(\theta)
     \exp(\mathrm i\xi x),
\end{equation}
which satisfies
\begin{equation}\label{eq: CIS_source}
    \left(\tau_{\mathrm C}^{-1}+\mathrm i\xi v\mu\right)\hat e^{\,n+1}=\frac{C_{p,\omega}}{2\pi}
    \left[
    \frac{\hat T_{\mathrm R}^{\,n}}{\tau_{\mathrm R}}
    +\frac{\hat T_{\mathrm N}^{\,n}}{\tau_{\mathrm N}}
    +\frac{T_0k}{\omega\tau_{\mathrm N}}
    \left(\mu\hat u_x^{\,n}+\zeta\hat u_y^{\,n}\right)
    \right].
\end{equation}
Hence, $\hat e^{\,n+1}$ is expressible in $\bm \hat{M}^n$ as
\begin{equation}\label{eq: e_solved}
    \hat e^{\,n+1}
    =\frac{1}
    {\tau_{\mathrm C}^{-1}+\mathrm i\xi v\mu}\frac{C_{p,\omega}}{2\pi}\left[\dfrac{\hat T_{\mathrm R}^{\,n}}{\tau_{\mathrm R}}
    +\dfrac{\hat T_{\mathrm N}^{\,n}}{\tau_{\mathrm N}}
    +\dfrac{T_0k}{\omega\tau_{\mathrm N}}
    \left(\mu\hat u_x^{\,n}+\zeta\hat u_y^{\,n}\right)\right].
\end{equation}
Taking the moments in \eqref{eq: T_loc_R}--\eqref{eq: drift_vel} gives
\begin{subequations}\label{eq: rows_derivation}
\begin{align}
    \hat T_{\mathrm R}^{\,n+1} &= \frac{1}{C_{\tau_{\mathrm{R}}}}\left\langle\frac{C_{p,\omega}}{\tau_{\mathrm R}}
    \left[\mathcal I^0_{p,\omega}\left(\frac{\hat T_{\mathrm R}^{\,n}}{\tau_{\mathrm R}} + \frac{\hat T_{\mathrm N}^{\,n}}{\tau_{\mathrm N}}\right)+\frac{T_0k_{p,\omega}\mathcal I^1_{p,\omega}}{\omega\tau_{\mathrm N}}\hat u_x^{\,n}\right]\right\rangle_{p}, \label{eq: TR_row_derivation}\\
    \hat T_{\mathrm N}^{\,n+1} &= \frac{1}{C_{\tau_{\mathrm N}}}\left\langle\frac{C_{p,\omega}}{\tau_{\mathrm N}}\left[
    \mathcal I^0_{p,\omega}\left(\frac{\hat T_{\mathrm R}^{\,n}}{\tau_{\mathrm R}}+\frac{\hat T_{\mathrm N}^{\,n}}{\tau_{\mathrm N}}\right)+\frac{T_0k_{p,\omega}\mathcal I^1_{p,\omega}}{\omega\tau_{\mathrm N}}\hat u_x^{\,n}\right]\right\rangle_{p}, \label{eq: TN_row_derivation}\\
    \hat u_x^{\,n+1} &= \frac{d}{T_0C_{\tau_{\mathrm N}}^1}\left\langle \frac{k_{p,\omega}C_{p,\omega}}{\omega\tau_{\mathrm N}}\left[\mathcal I^1_{p,\omega}\left(\frac{\hat T_{\mathrm R}^{\,n}}{\tau_{\mathrm R}}+\frac{\hat T_{\mathrm N}^{\,n}}{\tau_{\mathrm N}}\right)+\frac{T_0k_{p,\omega}\mathcal I^2_{p,\omega}}{\omega\tau_{\mathrm N}}\hat u_x^{\,n}\right]\right\rangle_{p}\label{eq: ux_row_derivation}\\
    \hat u_y^{\,n+1} &= \frac{d}{C_{\tau_{\mathrm N}}^1}\left\langle \frac{k_{p,\omega}^2C_{p,\omega}\mathcal J_{p,\omega}}{\omega^2\tau_{\mathrm N}^2}\right\rangle_{p}\hat u_y^{\,n}, \label{eq: uy_row_derivation}
\end{align}
\end{subequations}
with
\begin{subequations}\label{eq: resolvent_def}
\begin{align}
    \mathcal I_{p,\omega}^{\,j}(\xi) &:= \frac{1}{2\pi}\int_0^{2\pi} \frac{\mu^j}{\tau_{\mathrm C}^{-1}+\mathrm i\xi v_{p,\omega}\mu}\,\dd\theta, \qquad j=0,1,2, \\
    \mathcal J_{p,\omega}(\xi) &:= \frac{1}{2\pi}\int_0^{2\pi} \frac{\zeta^2}{\tau_{\mathrm C}^{-1}+\mathrm i\xi v_{p,\omega}\mu}\,\dd\theta,
\end{align}
\end{subequations}
which can be analytically calculated as shown in Appendix \ref{appendix: analytical_integrals}. Write in vector notation,
\begin{equation}\label{eq: G_CIS}
    \bm\hat{M}^{n+1}=\bm G_{\mathrm{CIS}}(\xi)\bm\hat{M}^n,
    \qquad
    \bm G_{\mathrm{CIS}}=
    \begin{pmatrix}
        G_{11}&G_{12}&G_{13}&0\\
        G_{21}&G_{22}&G_{23}&0\\
        G_{31}&G_{32}&G_{33}&0\\
        0&0&0&G_{44}
    \end{pmatrix},
\end{equation}
where the entries of the contraction matrix $[\bm G_{\mathrm{CIS}}]_{ij}$ are given in Appendix \ref{appendix: contraction_matrix_entries}. 

\paragraph{Contraction matrix of GSIS}

For GSIS, it first performs the kinetic intermediate step, thus, all the relations presented in the previous section hold by replacing the supscript $n+1$ with $n+1/2$, say,
\begin{equation}\label{eq: kinetic_half_step_map}
    \hat{\bm M}^{n+1/2}=\bm G_{\mathrm{CIS}}(\xi)\hat{\bm M}^n.
\end{equation}
Then the macroscopic synthetic equations are solved for $\hat{\bm M}^{n+1}$, where $\hat{\bm M}^{n+1/2}$ only goes into the high-order terms. To express the error amplitudes associated with the high-order terms in terms of $\hat{\bm M}^{n+1/2}$, let $\hat q_x^{\,n+1/2}$, $\hat\Pi_{xj}^{\,n+1/2}$, and $\hat p_{\mathrm R,j}^{\,n+1/2}$ ($j=x,y$) be the Fourier amplitudes of the heat flux, momentum-flux tensor, and resistive momentum and define differences in $\hat{\bm{M}}$ during the intermediate step  
\begin{equation}\label{eq: residual_def}
    \bm r := \left(\bm G_{\mathrm{CIS}}-\bm I\right)\hat{\bm M}^n = \hat{\bm M}^{n+1/2} - \hat{\bm M}^n = \left(r_T^{\mathrm R},r_T^{\mathrm N},r_u^x,r_u^y\right)^\top.
\end{equation}
Then, $\hat q_x^{\,n+1/2}$ satisfies
\begin{equation}\label{eq: en_id}
     \mathrm i\xi\hat q_x^{\,n+1/2}
     =-C_{\tau_{\mathrm R}}r_T^{\mathrm R}
      -C_{\tau_{\mathrm N}}r_T^{\mathrm N},
\end{equation}
which is obtained by taking $\langle\cdot\rangle_c$ to Eq. \eqref{eq: CIS_source}, while $\hat\Pi_{xj}^{\,n+1/2}$ and $\hat p_{\mathrm R,j}^{\,n+1/2}$ satisfy
\begin{equation}\label{eq: mom_id}
     \mathrm i\xi\hat\Pi_{xj}^{\,n+1/2}
     +\hat p_{\mathrm R,j}^{\,n+1/2}
     =-\frac{T_0C_{\tau_{\mathrm N}}^1}{d}\,r_u^j,
\end{equation}
which is obtained by multiplying Eq. \eqref{eq: CIS_source} by $k\zeta/\omega$ and taking $\langle\cdot\rangle_c$ to the resulting equation.

Finally, we define error-amplitude differences in solving the synthetic macroscopic equations,
\begin{equation}\label{eq: synthetic_increment_variables}
    \bm\delta := \left(\delta\hat T_{\mathrm N},\delta\hat u_x, \delta\hat u_y\right)^\top,
    \qquad
    \delta(\cdot):=(\cdot)^{n+1}-(\cdot)^{n+1/2}.
\end{equation}
On substituting the Fourier's error modes into Eqs. \eqref{eq: synth_energy_momentum} and using relations \eqref{eq: en_id} and \eqref{eq: mom_id}, we obtain the relation
\begin{equation}\label{eq: synthetic_increment_system}
    \bm S(\xi)\bm\delta=\bm R\bm r,
\end{equation}
where
\begin{equation}\label{eq: S_def}
    \bm S(\xi)=
    \begin{pmatrix}
        \kappa_{\mathrm N}\xi^2&\frac{T_0C_{vk}}{d}\mathrm i\xi&0\\[2pt]
        \left(\frac{C_{vk}}{d}-\eta\right)\mathrm i\xi&\left(3\mu_{\mathrm N}^1-\mu_{\mathrm N}^2\right)\xi^2+\frac{T_0C_{\tau_{\mathrm R}}^1}{d}&0\\[2pt]
        0&0&\mu_{\mathrm N}^1\xi^2+\frac{T_0C_{\tau_{\mathrm R}}^1}{d}
    \end{pmatrix},
    \quad
    \bm R=
    \begin{pmatrix}
        C_{\tau_{\mathrm R}}&C_{\tau_{\mathrm N}}&0&0\\[2pt]
        0&0&\dfrac{T_0C_{\tau_{\mathrm N}}^1}{d}&0\\[5pt]
        0&0&0&\dfrac{T_0C_{\tau_{\mathrm N}}^1}{d}
    \end{pmatrix}.
\end{equation}
The algebraic relation \eqref{eq: update_T_R} and the definition of $\mathrm{HoT}_{T^{\mathrm{loc}}_{\mathrm{R}}}^{n+1/2}$ give
$\delta\hat T_{\mathrm R}=\delta\hat T_{\mathrm N}
+\gamma\mathrm i\xi\delta\hat u_x$. Hence, the error amplification given by one macroscopic update becomes
\begin{equation}
    \hat{\bm M}^{n+1} = \hat{\bm M}^{n+1/2}+\bm L\bm\delta, \qquad 
    \bm L(\xi)=
    \begin{pmatrix}
        1&\gamma\mathrm i\xi&0\\
        1&0&0\\
        0&1&0\\
        0&0&1
    \end{pmatrix}.
\end{equation}
Combining \eqref{eq: kinetic_half_step_map}, \eqref{eq: residual_def}, and \eqref{eq: synthetic_increment_system} yields
\begin{equation}\label{eq: GH}
    \hat{\bm M}^{n+1} = \bm G_{\mathrm{GSIS}}(\xi)\hat{\bm M}^n,\qquad\bm G_{\mathrm{GSIS}}(\xi) = \bm G_{\mathrm{CIS}}(\xi) + \bm L(\xi)\bm S^{-1}(\xi)\bm R\left(\bm G_{\mathrm{CIS}}(\xi)-\bm I\right).
\end{equation}
If $\bm r = \bm 0$, then $\bm\delta=0$, so the synthetic solution preserves the fixed point of the kinetic solution and changes only the error-propagation spectrum.

\subsection{CIS dragging and GSIS acceleration}

\paragraph{CIS contraction factor.}

The contraction factors of the iterative schemes are defined as the maximum absolute eigenvalue of the contraction matrix, namely,
\begin{equation}
    \rho_{\mathrm{CIS}}(\xi) := \max_j\left|\lambda_j\left(\bm G_{\mathrm{CIS}}(\xi)\right)\right|, \qquad
    \rho_{\mathrm{GSIS}}(\xi) := \max_j\left|\lambda_j\left(\bm G_{\mathrm{GSIS}}(\xi)\right)\right|,
\end{equation}
To characterize the long-wavelength behavior of CIS, we consider the limit
$\xi=2\pi/L\to 0$. Note that
\begin{equation*}
    \bm G_{\mathrm{CIS}}(-\xi) = \bm D\,\bm G_{\mathrm{CIS}}(\xi)\,\bm D^{-1},
    \qquad  \bm D=\operatorname{diag}(1,1,-1,1),
\end{equation*}
which implies $\bm G_{\mathrm{CIS}}(-\xi)$ and $\bm G_{\mathrm{CIS}}(\xi)$ are similar and share the same eigenvalue spectrum. Hence, the spectral radius, namely the contraction factor of $\bm G_{\mathrm{CIS}}$ is an even function of $\xi$. In addition, for the limiting case $\xi = 0$, we have
\begin{equation*}
    \bm G_{\mathrm{CIS}}(0)(1,1,0,0)^{\top} = (1,1,0,0)^{\top},
\end{equation*}
thus, $\bm G_{\mathrm{CIS}}(0)$ has an eigenvalue equal to $1$. Since the remaining eigenvalues lie strictly inside the unit disc, the contraction factor can be written as
\begin{equation}\label{eq: CIS_contra_factor}
    \rho_{\mathrm{CIS}}(\xi) = 1 - c(\xi\bar\ell)^2 + O\!\left((\xi\bar\ell)^4\right), \qquad \xi \to 0,
\end{equation}
where $c > 0$ is a dimensionless coefficient determined by the material properties, and
\begin{equation}\label{eq: mean_collision_length}
    \bar\ell
    :=\frac{\sum_p\int C_{p,\omega}\ell_{p,\omega}\,\dd\omega}
    {\sum_p\int C_{p,\omega}\,\dd\omega},
\end{equation}
is the heat-capacity-weighted mean relaxation length. For the mode $\xi=2\pi/L$, \eqref{eq: CIS_contra_factor} gives
\[
    1-\rho_{\mathrm{CIS}}(L)
    =
    4\pi^2c\left(\frac{\bar\ell}{L}\right)^2
    +O\!\left(\left(\frac{\bar\ell}{L}\right)^4\right).
\]
Thus, $\rho_{\mathrm{CIS}}\to1$ as $L$ increases, and
reducing this long-wavelength error by a fixed factor requires
$O(L^2)$ iterations. This explains the deterioration of CIS
convergence at large characteristic lengths.

\paragraph{GSIS contraction factor.}

GSIS modifies this slow error component by adding synthetic correction in \eqref{eq: GH}. Let $\bm w(\xi)$ denote the eigenvector corresponding to $\rho(\xi)$, we have
\begin{equation*}
    \bm r_{\xi} := \left(\bm G_{\mathrm{CIS}}-\bm I\right)\bm w = O(\xi^2).
\end{equation*}
More specifically, $\bm w(\xi)$ has temperature components of order $O(1)$ and longitudinal drift component of order $O(\xi)$, so
\begin{equation*}
    \bm R\bm r_{\xi} =
    \begin{pmatrix}
        O(\xi^2)\\
        O(\xi^3)\\
        0
    \end{pmatrix},
\end{equation*}
and consequently,
\begin{equation}\label{eq: Fourier_GSIS_correction_order}
    \bm S^{-1}\bm R\bm r_{\xi} =
    \begin{pmatrix}
        O(1)\\
        O(\xi)\\
        0
    \end{pmatrix},
    \qquad
    \bm L(\xi)\bm S^{-1}(\xi)\bm R \left(\bm G_{\mathrm{CIS}}(\xi)-\bm I\right)\bm w(\xi) = O(1).
\end{equation}
Thus, although CIS changes the long-wavelength energy error only by $O(\xi^2)$ per iteration, \eqref{eq: Fourier_GSIS_correction_order} converts this residual into a finite leading-order correction, which can therefore shift the long-wavelength amplification factor away from the CIS limit of unity rather than merely perturbing it by another $O(\xi^2)$ term. This leads to
\begin{equation*}
    \limsup_{\xi\to0}\rho_{\mathrm{GSIS}}(\xi)<1,
\end{equation*}
which gives an effective acceleration. The numerical evaluation below verifies this behavior for a graphene model.

\subsection{Numerical evaluation}\label{sec: fourier_num_evl}

We numerically evaluate the contraction factors of CIS and GSIS using a suspended-monolayer-graphene model \cite{guo2017heat}, which contains, the in-plane longitudinal (LA), transverse acoustic (TA) and the out-of-plane flexural acoustic (ZA) phonon polarization branches. The integral over phonon spectrum, $\langle \cdot \rangle_{p}$, is approximated by the same eight-node Gauss--Legendre rule on $[0,\omega_{\max}^{p}]$ used in the benchmark calculations in \Cref{sec: numerical_validation}. The angular $\theta$ dependence is evaluated with the analytical closed-form resolvents in \Cref{appendix: analytical_integrals}. Thus, the results below correspond to the frequency-discrete non-gray model with an angularly continuous Fourier symbol.

\begin{figure}[pos=htbp]
\centering
\subfloat[spectral radius w.r.t $\xi = 2\pi/L$.]{
  \includegraphics[width=0.47\textwidth]{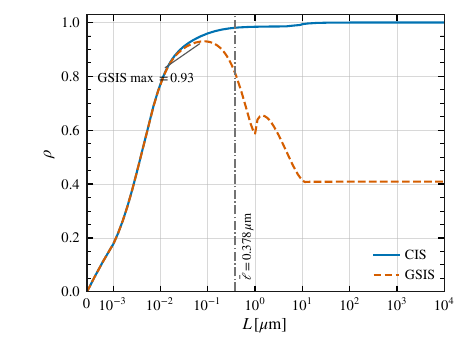}
  \label{fig: spectral_radius_a}}
\hfill
\subfloat[Implied iteration count.]{
  \includegraphics[width=0.47\textwidth]{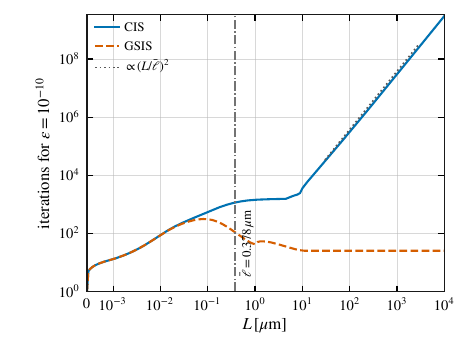}
  \label{fig: spectral_radius_b}}
\caption{Fourier convergence factors for the frequency-discretized graphene model at $T_0=300\,\mathrm K$. (a) Spectral radius $\rho(2\pi/L)$ for CIS and GSIS as the physical characteristic length $L$ is varied. The dashed vertical line marks the heat-capacity-weighted mean combined relaxation length $\bar\ell=0.378\,\mu\mathrm m$ and is shown only as a reference spectral length, not as a regime boundary. (b) Implied number of iterations $N=\lceil\ln\varepsilon/\ln\rho\rceil$ for $\varepsilon:=\|\bm{M}^{n+1}-\bm{M}^{n}\|/\|\bm{M}^{1}-\bm{M}^{0}\|=10^{-10}$. The CIS count grows quadratically at large $L$, whereas the GSIS count remains bounded and approaches approximately $26$.}
\label{fig: spectral_radius}
\end{figure}

At $T_0=300\,\mathrm K$, the heat-capacity-weighted mean relaxation length \eqref{eq: mean_collision_length} is
$\bar\ell=0.378\,\mu\mathrm m$. It is retained only as a reference spectral length; the scale variable in the evaluation is the physical characteristic length $L$. The coefficients entering the contraction matrices are listed in \Cref{tab: fourier_constants}. \Cref{fig: spectral_radius} illustrate the contraction factors of GSIS and CIS, covering $L\in(0,10^4]~\mu\mathrm m$. Representative values are listed in \Cref{tab: spectral_radius}.

\begin{table}[pos=htbp]
\centering
\small
\begin{tabular}{lll}
\hline
$C_V=3.426\times10^{-4}$
&$C_{\tau_{\mathrm R}}=1.610\times10^{7}$
&$C_{\tau_{\mathrm N}}=4.837\times10^{7}$\\
$C_{\tau_{\mathrm N}}^{1}=1.463\times10^{1}$
&$C_{\tau_{\mathrm R}}^{1}=3.633\times10^{-1}$
&$C_{vk}=5.273\times10^{-4}$\\
$C_k=1.508\times10^{-11}$
&$\kappa_{\mathrm N}=1.250\times10^{-6}$
&$\eta=1.901\times10^{-4}$\\
$\mu_{\mathrm N}^{1}=7.207\times10^{-13}$
&$\mu_{\mathrm N}^{2}=1.712\times10^{-12}$
&$\gamma=-4.204\times10^{-10}$\\
\hline
\end{tabular}
\caption{Coefficients of the frequency-discrete non-gray graphene model at $T_0=300\,\mathrm K$, in their corresponding SI units. Each branch uses eight Gauss--Legendre nodes for integral with respect to phonon frequency.}
\label{tab: fourier_constants}
\end{table}

\begin{table}[pos=htbp]
\centering
\small
\begin{tabular}{rccrrr}
\hline
$L\,[\mu\mathrm m]$
&$\rho_{\mathrm{CIS}}$
&$\rho_{\mathrm{GSIS}}$
&$N_{\mathrm{CIS}}$
&$N_{\mathrm{GSIS}}$
&speed-up\\
\hline
$10^{-3}$&$0.1770$&$0.1769$&$14$&$14$&$1$\\
$10^{-1}$&$0.9596$&$0.9290$&$558$&$313$&$1.8$\\
$1$&$0.9843$&$0.5876$&$1\,452$&$44$&$33$\\
$3$&$0.9851$&$0.5886$&$1\,537$&$44$&$35$\\
$10$&$0.9936$&$0.4143$&$3\,607$&$27$&$1.3\times10^2$\\
$30$&$0.9992$&$0.4088$&$3.07\times10^4$&$26$&$1.18\times10^3$\\
$10^2$&$0.9999$&$0.4089$&$3.11\times10^5$&$26$&$1.20\times10^4$\\
$10^3$&$1.0000$&$0.4089$&$2.97\times10^7$&$26$&$1.14\times10^6$\\
$10^4$&$1.0000$&$0.4089$&$2.97\times10^9$&$26$&$1.14\times10^8$\\
\hline
\end{tabular}
\caption{Contraction factors (w.r.t. $\xi = 2\pi / L$) and implied iteration counts for $\varepsilon:=\|\bm{M}^{n+1}-\bm{M}^{n}\|/\|\bm{M}^{1}-\bm{M}^{0}\|=10^{-10}$. The displayed spectral radii are rounded, whereas $N=\lceil\ln\varepsilon/\ln\rho\rceil$ is evaluated from the unrounded eigenvalues.}
\label{tab: spectral_radius}
\end{table}

\paragraph{Small characteristic lengths.}
As $L$ decreases, the mode-dependent ratios $\mathrm{Kn}_{p,\omega} = \ell_{p,\omega}/L$ increase. In the limit $L\to 0$, $\mathrm{Kn}_{p,\omega} \to\infty$ for every discrete phonon mode in the phase space. Then the evaluation of the the entries of $\bm G_{\mathrm{CIS}}$ implies that $\bm G_{\mathrm{CIS}} \to \bm 0$, so that the kinetic sweep therefore significantly damps the corresponding error modes while the additional synthetic correction becomes less significant. Consequently, the CIS and GSIS contraction factors approach each other at small characteristic lengths. Hence, at very small $L$, CIS is already efficient and GSIS provides little further correction.

\paragraph{Large characteristic lengths.}
As $L$ increases, $\mathrm{Kn}_{p,\omega}$ decreases for all phonons. Equation~\eqref{eq: CIS_contra_factor} then gives $\rho_{\mathrm{CIS}}(L) =1- O(L^{-2})$, so $\rho_{\mathrm{CIS}}\to 1 $ as $L\to\infty$. Numerically, $1-\rho_{\mathrm{CIS}}=7.4\times10^{-5}$, $7.7\times10^{-7}$, and $7.8\times10^{-9}$ at $L=10^2$, $10^3$, and $10^4\,\mu\mathrm m$, respectively, and the implied CIS iteration count consequently grows quadratically with $L$. In GSIS, by contrast, solving synthetic macroscopic transforms the $O(\xi^2)$ CIS residual of the long-wavelength mode into a finite leading-order correction, preventing the corresponding contraction factor from approaching unity. The GSIS spectral radius reaches a maximum of approximately $0.93$ near $L\simeq0.08,\mu\mathrm m$ and approaches $0.409$ as $L$ increases, corresponding to about $26$ iterations for $\varepsilon:=\|\bm{M}^{n+1}-\bm{M}^{n}\|/\|\bm{M}^{1}-\bm{M}^{0}\|=10^{-10}$ ($\bm{M}^{0}$ is initial data). The resulting speed-up is approximately $1.2\times10^4$ at $L=10^2\,\mu\mathrm m$ and $1.1\times10^6$ at $L=10^3\,\mu\mathrm m$.

\section{Asymptotic preserving analysis}
\label{sec: ap_analysis}

The Fourier analysis in \Cref{sec: fourier_analysis} concerns the contraction of iteration. We now consider a different question: after GSIS has converged, does the numerical solver for the kinetic system reduce to a consistent discretization to certain macroscopic equations when the phonon transport approaches the diffusion or hydrodynamic limit?  This is the asymptotic-preserving (AP) property considered in the following \cite{jin2022asymptotic,su2020fast,liu2022fast}. 

\subsection{Fixed-point compatibility}
\label{sec: ap_fixed_point}

We use subscript `$*$' to denote the solution from the macroscopic synthetic equations~\eqref{eq: synth_energy_momentum}, and subscript `$h$' for those by taking moments to the solution of the  kinetic equation~\eqref{eq: GSIS_kinetic_half_step}. The iteration index is omitted since we focus on the converged result, i.e. when the fixed point of iteration is reached. Similarly to \eqref{eq: synthetic_increment_variables}, the operator $\delta$ denotes the difference between them, i.e. $\delta \bm M = \bm M_{*} - \bm M_{h}$. The kinetic equation at the fixed point reads
\begin{equation}\label{eq: kinetic_fixed_point_eq}
    \bm v\cdot\nabla_{\bm x}e=\frac{e_{\mathrm R}^{\mathrm{eq}}\left(T_{\mathrm R,*}^{\mathrm{loc}}\right)-e}{\tau_{\mathrm R}}+\frac{e_{\mathrm N}^{\mathrm{eq}}\left(T_{\mathrm N,*}^{\mathrm{loc}},\bm u_{*}\right)-e}{\tau_{\mathrm N}}.
\end{equation}
Conservation of energy and quasi--momentum gives
\begin{subequations}\label{eq: increment_consti_rela}
\begin{align}
    \nabla \cdot \bm q_{h} &= C_{\tau_{\mathrm{R}}} \delta T_{\mathrm{R}} + C_{\tau_{\mathrm{N}}} \delta T_{\mathrm{N}}, \label{eq: increment_consti_rela_a} \\
    \nabla \cdot \bm \Pi_{h} &= - \bm p_{\mathrm{R},h} + \frac{T_0 C_{\tau_{\mathrm{N}}}^1}{d}\delta \bm u \label{eq: increment_consti_rela_b}
\end{align}
\end{subequations}
In addition, \eqref{eq: update_T_R} and \eqref{eq: HoT_T} yield
\begin{equation}\label{eq: increment_TR}
    \delta T_{\mathrm{R}} = \delta T_{\mathrm{N}} + \gamma \nabla \cdot \delta \bm u.
\end{equation}
Applying some algebraic calculations to the synthetic equations \eqref{eq: synth_energy_momentum}, the definition of high-order terms \eqref{eq: HoT_defs}, as well as Eqs. \eqref{eq: increment_consti_rela} and \eqref{eq: increment_TR}, we obtain the homogeneous linear increment system with respect to $\delta T_{\mathrm{N}}$ and $\delta \bm u$:
\begin{subequations}\label{eq: increment_homo_eq}
\begin{align}
    -\kappa_{\mathrm{N}} \Delta \delta T_{\mathrm{N}} + \left(C_{\tau_{\mathrm{R}}} + C_{\tau_{\mathrm{N}}}\right) \delta T_{\mathrm{N}} + \left(\frac{T_0C_{vk}}{d} + C_{\tau_{\mathrm{R}}} \gamma\right)\nabla \cdot \delta \bm u &= 0, \\
    \left(\frac{C_{vk}}{d} - \eta\right) \nabla \delta T_{\mathrm{N}} - \mu^1_{\mathrm{N}}\left[\Delta \delta \bm u + 2 \nabla\left(\nabla \cdot \delta \bm u\right)\right] + \mu^2_{\mathrm{N}} \nabla\left(\nabla \cdot \delta \bm u\right) + \left(\frac{T_0 \left(C_{\tau_{\mathrm{R}}}^1 + C_{\tau_{\mathrm{N}}}^1\right)}{d}\right)\delta \bm u &= 0.
\end{align}
\end{subequations}
In other words, at a fixed point in GSIS, the differences between the synthetic fields and the moments of the kinetic solution satisfy this homogeneous linear increment problem \eqref{eq: increment_homo_eq}. We assume that this fixed-point increment problem is nonsingular for the admissible material parameters and boundary setting considered here, then the homogeneous problem admits a unique solution, i.e. the trivial solution
\begin{equation}
    \delta T_{\mathrm{N}} = 0, \qquad \delta \bm u = 0.
\end{equation}
The reconstruction relation then also gives $\delta T_{\mathrm{R}}=0$. Hence, the converged synthetic macroscopic fields coincide with the corresponding moments of the converged kinetic solution.

The following analysis is performed under the fixed--point compatibility, i.e. $\bm M_{*} = \bm M_{h}$. We will claim that with certain requirements to spatial discretization for solving the kinetic equation~\eqref{eq: GSIS_kinetic_half_step}, the converged solution $\bm M_{*}$ is that of a Guyer–Krumhansl-like equation at the hydrodynamic limit or of Fourier's law of heat conduction at the diffusive limit, respectively.

\subsection{Normal-scattering-dominated hydrodynamic limit}
\label{sec: ap_hydrodynamic}
To study the asymptotic behavior of the macroscopic synthetic equations at the hydrodynamic limit, we use the same scaling in \Cref{sec: gsis_ce_analysis}, and the exact same Chapman-Enskog analysis holds. Then, if the order of a spatial discretization, say $K$, to the kinetic equation~\eqref{eq: GSIS_kinetic_half_step} satisfies $K\geq2\alpha$ with $\alpha>1$, the high-order terms entering the synthetic system are all at least $O(\epsilon^2)$ with cell size $\Delta x$ not resolving the small scale, i.e., $\Delta x/L\sim O(\epsilon^{1/\alpha})$. Thus, the synthetic heat flux \eqref{eq: update_q_mac} yields
\begin{equation}\label{eq: ap_hydro_u2q}
    \bm u = \frac{d}{T_{0}C_{vk}} \bm q + \frac{d \kappa_{\mathrm{N}}}{T_{0}C_{vk}} \nabla_{\bm x} \vartheta_{\mathrm{N}} + O(\epsilon^2),
\end{equation}
where $\vartheta_{\mathrm{N}}:=T^{\mathrm{loc}}_{\mathrm{N}} - T_0$ in shorthand. This derives
\begin{subequations}\label{eq: ap_hydro_du}
\begin{align}
    \nabla_{\bm x} \cdot \bm u &= \frac{d}{T_{0}C_{vk}} \nabla_{\bm x} \cdot \bm q + \frac{d \kappa_{\mathrm{N}}}{T_{0}C_{vk}} \Delta_{\bm x} \vartheta_{\mathrm{N}} + O(\epsilon^2) \\
    \Delta_{\bm x} \bm u &= \frac{d}{T_{0}C_{vk}} \Delta_{\bm x} \bm q + \frac{d \kappa_{\mathrm{N}}}{T_{0}C_{vk}} \nabla_{\bm x} \Delta_{\bm x} \vartheta_{\mathrm{N}} + O(\epsilon^2), \\
    \nabla_{\bm x}\left(\nabla_{\bm x} \cdot \bm u\right) &= \frac{d}{T_{0}C_{vk}} \nabla_{\bm x} \left( \nabla_{\bm x} \cdot \bm q \right) + \frac{d \kappa_{\mathrm{N}}}{T_{0}C_{vk}} \nabla_{\bm x} \Delta_{\bm x} \vartheta_{\mathrm{N}} + O(\epsilon^2).
\end{align}
\end{subequations}
Substituting \eqref{eq: ap_hydro_du} into \eqref{eq: synth_energy_momentum}, we have
\begin{subequations}\label{eq: ap_hydro_GK_0}
\begin{align}
    \nabla_{\bm x} \cdot \bm q &= 0 + O(\epsilon^2), \\
    \frac{C_{\tau_{\mathrm{R}}}^1}{C_{vk}} \bm q - \left(\frac{C_{vk}}{d} - \eta\right)\nabla_{\bm{x}} \vartheta_{\mathrm{N}} - \frac{\mu_{\mathrm{N}}^1 d}{T_0 C_{vk}}\Delta_{\bm{x}} \bm q -\left(\frac{2\mu_{\mathrm{N}}^1 d }{T_0 C_{vk}} -  \frac{\mu_{\mathrm{N}}^2 d \kappa_{\mathrm{N}}}{T_0 C_{vk}} \right)\nabla_{\bm{x}}\left(\nabla_{\bm{x}} \cdot \bm q \right) +& \nonumber \\
    \left\{-\frac{2\mu_{\mathrm{N}}^1 d \kappa_{\mathrm{N}}}{T_0 C_{vk}} \nabla_{\bm x} \Delta_{\bm x} \vartheta_{\mathrm{N}} + \frac{\mu_{\mathrm{N}}^2 d \kappa_{\mathrm{N}}}{T_0 C_{vk}} \nabla_{\bm x} \Delta_{\bm x} \vartheta_{\mathrm{N}} + \frac{C_{\tau_{\mathrm{R}}}^1 \kappa_{\mathrm{N}}}{C_{vk}} \nabla_{\bm x} \vartheta_{\mathrm{N}} \right\}  &= 0 + O(\epsilon^2)
\end{align}
\end{subequations}
The definition of the transport coefficients in \Cref{sec: A1} gives $\mu_{\mathrm{N}}^1 = O(\epsilon)$, $\mu_{\mathrm{N}}^2 = O(\epsilon)$, $\kappa_{\mathrm{N}} = O(\epsilon)$, $C_{\tau_{\mathrm{R}}}^1 = O(\epsilon)$, which implies that
\begin{equation*}
    \frac{2\mu_{\mathrm{N}}^1 d \kappa_{\mathrm{N}}}{T_0 C_{vk}} \nabla_{\bm x} \Delta_{\bm x} \vartheta_{\mathrm{N}} = O(\epsilon^2),
    \qquad
    \frac{\mu_{\mathrm{N}}^2 d \kappa_{\mathrm{N}}}{T_0 C_{vk}} \nabla_{\bm x} \Delta_{\bm x} \vartheta_{\mathrm{N}} = O(\epsilon^2),
    \qquad
    \frac{C_{\tau_{\mathrm{R}}}^1 \kappa_{\mathrm{N}}}{C_{vk}} \nabla_{\bm x} \vartheta_{\mathrm{N}} = O(\epsilon^2).
\end{equation*}
Hence, \eqref{eq: ap_hydro_GK_0} further reduces to
\begin{subequations}\label{eq: ap_hydro_GK}
\begin{align}
    \nabla_{\bm{x}} \cdot \bm q &= 0 + O(\epsilon^2), \\
    \frac{C_{\tau_{\mathrm{R}}}^1}{C_{vk}} \bm q - \left(\frac{C_{vk}}{d} - \eta\right)\nabla_{\bm{x}} \vartheta_{\mathrm{N}} - \frac{\mu_{\mathrm{N}}^1 d}{T_0 C_{vk}}\Delta_{\bm{x}} \bm q -d\left(\frac{2\mu_{\mathrm{N}}^1}{T_0 C_{vk}} -  \frac{\mu_{\mathrm{N}}^2 \kappa_{\mathrm{N}}}{T_0 C_{vk}} \right)\nabla_{\bm{x}}\left(\nabla_{\bm{x}} \cdot \bm q \right) &= 0 + O(\epsilon^2),
\end{align}
\end{subequations}
which proves that the macroscopic synthetic equation \eqref{eq: synth_energy_momentum} asymptotically converges to a Guyer-Krumhansl-like equation~\cite{Shang2020HydroPhonon} with respect to $\bm q$ and $T_{\mathrm{N}}$ for stationary problems as $\epsilon\ll1$ under the hydrodynamic limit scaling.


\subsection{Resistive-scattering-dominated diffusion limit}
\label{sec: ap_diffusion}
For the diffusive limit where the resistive scattering dominates, to allow the relaxation time $\tau_{\mathrm{R}}$ to become small, we introduce the small parameter $\epsilon$ into the kinetic equation as   
\begin{equation}\label{eq: ap_diff_kinetic}
    \bm{v}\cdot\nabla_{\bm{x}} e = \frac{e_{\mathrm{R}}^{\mathrm{eq}} - e}{\epsilon\widehat{\tau}_{\mathrm{R}}} + \frac{e_{\mathrm{N}}^{\mathrm{eq}} - e}{\tau_{\mathrm{N}}},
\end{equation}
where $\tau_{\mathrm{R}} = \epsilon \widehat{\tau}_{\mathrm{R}}$.
Substituting expansion~\eqref{eq: CE_expansion} and collecting equal powers of $\epsilon$, we could solve that
\begin{subequations}
\begin{align}
    e^{(0)} &= \frac{C_{p,\omega}}{\mathrm B}\vartheta_{\mathrm R} \label{eq: ap_diff_e0} \\
    e^{(1)} &= - \widehat{\tau}_{\mathrm{R}} \bm v \cdot \nabla_{\bm x} e^{(0)} + \frac{\widehat{\tau}_{\mathrm{R}}}{\tau_{\mathrm{N}}}\frac{C_{p,\omega}}{\mathrm{B}}\vartheta_{\mathrm{R}}, \label{eq: ap_diff_e1}
\end{align}
\end{subequations}
with $\vartheta_{\mathrm{R}}:=T^{\mathrm{loc}}_{\mathrm{R}} - T_0$. Therefore, the heat flux evaluated by the expansion becomes
\begin{equation}\label{eq: ap_diff_q_expansion}
    \bm q = \left\langle \bm v e \right\rangle_{c} = -\epsilon\frac{1}{d}\left\langle \widehat{\tau}_{\mathrm{R}} |\bm v|^2 C_{p,\omega}\right\rangle_{p} \nabla_{\bm x} \vartheta_{\mathrm{R}} + O(\epsilon^2) =: - \kappa_{\mathrm{R}} \nabla_{\bm x} \vartheta_{\mathrm{R}} + O(\epsilon^2), \qquad \kappa_{\mathrm{R}} = \frac{1}{d}\left\langle \tau_{\mathrm{R}} |\bm v|^2 C_{p,\omega}\right\rangle_{p}.
\end{equation}
Substituting this into \eqref{eq: HoT_q} gives
\begin{equation}\label{eq: ap_diff_hot_q}
    \mathrm{HoT}_{\bm q} = - \kappa_{\mathrm{R}} \nabla_{\bm x} \vartheta_{\mathrm{R}} - \frac{T_0 C_{vk}}{d} \bm u + \kappa_{N} \nabla_{\bm x} \vartheta_{\mathrm{N}} + O(\epsilon^2).
\end{equation}
Combining \eqref{eq: ap_diff_hot_q} and the macroscopic synthetic equation \eqref{eq: synth_energy_momentum_a} yields
\begin{equation}\label{eq: ap_diff_q_fourier}
    \nabla \cdot \bm q = 0 + O(\epsilon^2),\qquad \bm{q}=- \kappa_{\mathrm{R}} \nabla_{\bm x} \vartheta_{\mathrm{R}} + O(\epsilon^2).
\end{equation}
Therefore, the macroscopic equation \eqref{eq: synth_energy_momentum_a} degenerates into Fourier's law of heat conduction, when $\epsilon\ll1$.

Furthermore, we show that the other macroscopic synthetic equation \eqref{eq: synth_energy_momentum_b} does not offer any additional equation under diffusive scaling. Following similar paths in deriving \eqref{eq: ap_diff_hot_q}, we have
\begin{subequations}\label{eq: ap_diff_hot_pi_pr}
\begin{align}
    \nabla \cdot \mathrm{HoT}_{\bm \Pi} &= \mu_{\mathrm{N}}^1 \left[\Delta_{\bm x} \bm u + 2 \nabla_{\bm x}\left(\nabla_{\bm x} \cdot \bm u\right)\right] - \mu_{\mathrm{N}}^2 \nabla_{\bm x}\left(\nabla_{\bm x} \cdot \bm u\right) + O(\epsilon^2) \\
    \mathrm{HoT}_{\bm p_{\mathrm{R}}} &= -\left(\frac{C_{vk}}{d} \nabla - \eta \right)\nabla_{\bm{x}}\vartheta_{\mathrm{N}} - \frac{T_0C_{\tau_{\mathrm{R}}}^1}{d} \bm u + O(\epsilon^2).
\end{align}
\end{subequations}
Substituting \eqref{eq: ap_diff_hot_pi_pr} into \eqref{eq: synth_energy_momentum_b} implies the trivial identity
\begin{equation}
    0 = O(\epsilon^2).
\end{equation}

Therefore, at the diffusive limit, the synthetic energy equation is asymptotically reduced to Fourier’s heat-conduction equation \eqref{eq: ap_diff_hot_q} and \eqref{eq: ap_diff_q_expansion}, while the quasi-momentum equation becomes asymptotically redundant. Hence, the macroscopic synthetic system consistently recovers the Fourier limit as $\epsilon \ll 1$.

\section{DG and HDG Spatial Discretization}
\label{sec: dg_hdg}

For numerical implementation, the kinetic equation is spatially discretized by the nodal discontinuous Galerkin (DG) method~\cite{NodalDG} to provide a high-order approximation and meet the requirement for the AP property, see the description in \Cref{sec: ap_hydrodynamic}, while the macroscopic synthetic equation is solved by the hybridizable discontinuous Galerkin (HDG) method~\cite{Sevilla2018HDG}. Furthermore, finite-element discretization has the advantage of flexibly treating complex geometries. 

\subsection{DG solution of the kinetic equation}
\label{sec: DG_scheme}

For a simple presentation, the stationary kinetic equation~\eqref{eq: GSIS_kinetic_half_step} is rewritten as
\begin{equation}\label{eq: DG_kinetic}
    \frac{1}{\tau_{\mathrm C}}e^{n+1/2}+\bm v\cdot\nabla_{\bm x}e^{n+1/2}=S^n,
    \qquad
    S^n:=\frac{e_{\mathrm R}^{\mathrm{eq}}(T_{\mathrm R}^{\mathrm{loc},n})}{\tau_{\mathrm R}}
        +\frac{e_{\mathrm N}^{\mathrm{eq}}(T_{\mathrm N}^{\mathrm{loc},n},\bm u^n)}{\tau_{\mathrm N}}.
\end{equation}
Let $\mathcal T_h=\{\Delta_l\}_{l=1}^{N_{\mathrm{el}}}$ be a conforming partition of a domain $\Delta\subset\mathbb R^d$ consisting of $N_{\mathrm{el}}$ $d$-dimensional simplexes, and let $\Upsilon=\bigcup_{l=1}^{N_{\mathrm{el}}}\partial \Delta_l := \bigcup_{c=1}^{N_{\mathrm{fc}}}\Gamma_c$ be the mesh skeleton consisting of $N_{\mathrm{fc}}$ $(d-1)$-dimensional cell faces. We introduce the following discrete functional spaces
\begin{equation}\label{eq: DG_space}
    \mathcal{V}_h:=\left\{\psi\in L^2(\Delta):
    \psi|_{\Delta_l}\in\mathbb P_K(\Delta_l),\quad l=1,\ldots,N_{\mathrm{el}}\right\},
\end{equation}
where $\mathbb P_K(\Delta_l)$ stands for the space of polynomials of complete degree at most $K$ in $\Delta_l$. In this work, we choose the nodal Lagrangian polynomials as its basis. The element-wise weak form of the equation with an upwind numerical flux reads
\begin{equation}\label{eq: DG_kinetic_weak_form}
    \frac{1}{\tau_{\mathrm C}}(\psi,e^{n+1/2})_{\Delta_l}
    -(\bm v\cdot\nabla_{\bm{x}}\psi,e^{n+1/2})_{\Delta_l}
    +\left\langle\psi,(\bm v\cdot\bm n)e^{n+1/2}\right\rangle_{\partial \Delta_l^{\mathrm{out}}}
    =(\psi,S^n)_{\Delta_l}
    -\left\langle\psi,(\bm v\cdot\bm n)e_{\mathrm{ext}}^{n+1/2}\right\rangle_{\partial \Delta_l^{\mathrm{in}}}.
\end{equation}
for all $\psi\in\mathcal{V}_h$, where $(\cdot,\cdot)_{\Delta_l}$ and $\langle\cdot,\cdot\rangle_{\partial\Delta_l}$ denote the inner product in the corresponding domain; $\partial \Delta_l^{\mathrm{in}}$/$\partial \Delta_l^{\mathrm{out}}$ indicate the inflow/outflow face; $\bm n$ is the outward unit normal vector of $\partial \Delta_l$; the subscript `$\mathrm{ext}$' denotes the upwind neighboring element or the physical boundary. 




\paragraph{Isothermal boundary.}
At a temperature $T_b$, the incoming distribution function is the equilibrium
\begin{equation}\label{eq: BC_iso}
    e_{\mathrm{ext}}
    =\frac{C_{p,\omega}}{\mathrm B}(T_b-T_0).
\end{equation}

\paragraph{Diffuse non-thermalizing adiabatic boundary.}
The incoming distribution function is defined as follows to ensure energy conservation.
\begin{equation}\label{eq: BC_adi}
    e_{\mathrm{ext}}
    =-\frac{\displaystyle
       \left\langle\bm (\bm v\cdot\bm n)e^{n+1/2}\right\rangle_{c,\ \bm v\cdot\bm n>0}}
      {\left\langle
       (\bm v\cdot\bm n)\right\rangle_{c,\ \bm v\cdot\bm n<0}}.
\end{equation}

\paragraph{Periodic heat-flux boundary.}
Let $\partial\Delta_L$ and $\partial\Delta_R$ be paired periodic faces with corresponding points $\bm x_L$ and $\bm x_R$, and let $\Delta T$ be a temperature increment. Periodicity of the non-equilibrium part of the distribution gives
\begin{subequations}\label{eq: BC_periodic}
\begin{align}
    e_{\mathrm{ext}}(\bm x_L) &= e(\bm x_R)+\frac{C_{p,\omega}}{\mathrm B}\Delta T, \qquad \bm v\cdot\bm n_L<0, \label{eq: BC_periodic_L}\\
    e_{\mathrm{ext}}(\bm x_R) &= e(\bm x_L)-\frac{C_{p,\omega}}{\mathrm B}\Delta T, \qquad \bm v\cdot\bm n_R<0. \label{eq: BC_periodic_R}
\end{align}
\end{subequations}

In CIS, the DG formulation is understood by replacing the iteration index `$n+1/2$' by `$n+1$'. 

\subsection{HDG solution of the macroscopic synthetic equations}

In HDG implementation, we solve the increments relative to the synthetic update from the intermediate kinetic state, say, $\delta T_{\mathrm{N}} := T_{\mathrm N}^{\mathrm{loc},n+1}-T_{\mathrm N}^{\mathrm{loc},n+1/2}$ and $\delta\bm u:= \bm u^{n+1}-\bm u^{n+1/2}$. In addition, introduce the auxiliary gradients $\bm g:=\nabla_{\bm x}\delta T_{\mathrm N}$, $\bm G:=\nabla_{\bm x}\delta\bm u$, the corresponding increments in the constitutive relations are
\begin{equation}\label{eq: HDG_increment_fluxes}
\bm f^\delta := -\kappa_{\mathrm N}\bm g +\frac{T_0C_{vk}}{d}\delta\bm u,\quad
\bm\sigma^\delta := \frac{C_{vk}}{d}\delta T_{\mathrm N}\bm I - \mu_{\mathrm N}^1 \left[ \bm G+\bm G^{\top} + \operatorname{tr}(\bm G)\bm I \right] + \mu_{\mathrm N}^2 \operatorname{tr}(\bm G)\bm I, \quad
\bm r^\delta := \frac{T_0C_{\tau_{\mathrm R}}^1}{d}\delta\bm u -\eta\bm g.
\end{equation}
Then, the synthetic equations \eqref{eq: synth_energy_momentum} are transformed into the following first-order system
\begin{subequations}\label{eq: HDG_increment_system}
\begin{align}
    \bm g-\nabla_{\bm x}\delta T_{\mathrm N}&=\bm0,
    \label{eq: HDG_sys_A1}\\
    \bm G-\nabla_{\bm x}\delta\bm u&=\bm0,
    \label{eq: HDG_sys_A2}\\
    \nabla_{\bm x}\cdot\bm f^\delta
    &=-\nabla_{\bm x}\cdot\bm q^{n+1/2},
    \label{eq: HDG_sys_C1}\\
    \nabla_{\bm{x}}\cdot\bm{\sigma}^{\delta}+\bm{r}^\delta &= -\nabla_{\bm{x}}\cdot\bm{\Pi}^{n+1/2} -\bm p_{\mathrm R}^{n+1/2},
    \label{eq: HDG_sys_C2}
\end{align}
\end{subequations}
Note that the right-hand side of this HDG problem is determined entirely by the kinetic solution $e^{n+1/2}$, treating as known source terms.

In addition to the functional space $\mathcal{V}_h$ in \eqref{eq: DG_space}, we also introduce a discrete space on the mesh skeleton, written as
\begin{equation}\label{eq: HDG_trace_space}
    \mathcal{W}_h:=\left\{\mu\in L^2(\Upsilon): \mu|_{\Gamma_c}\in\mathbb P_K(\Gamma_c),\quad c=1,\ldots,N_{\mathrm{fc}}\right\}.
\end{equation}
The HDG problem finds $\{\delta T_{\mathrm{N}},\delta\bm u,\bm g,\bm G\} \in \mathcal{V}_h \times [\mathcal{V}_h]^d \times [\mathcal{V}_h]^d \times [\mathcal{V}_h]^{d \times d}$ and the traces $\{\widehat{\delta T_{\mathrm{N}}},\widehat{\delta\bm u}\} \in \mathcal{W}_h \times [\mathcal{W}_h]^d$. Particularly, they are found from the following weak form for all $\{\psi,\bm{\phi},\bm{\Phi}\}\in \mathcal{V}_h\times[\mathcal{V}_h]^d\times[\mathcal{V}_h]^{d\times d}$
\begin{subequations}\label{eq: HDG_weak}
\begin{align}
(\bm\phi,\bm g)_{\Delta_l} + (\nabla_{\bm x}\cdot\bm\phi,\delta T_{\mathrm N})_{\Delta_l} &= \left\langle \bm\phi\cdot\bm n, \widehat{\delta T}_{\mathrm N} \right\rangle_{\partial \Delta_l}, \label{eq: HDG_weak_aux_T} \\
(\bm\Phi,\bm{G})_{\Delta_l} + \left(\nabla_{\bm x}\cdot\bm\Phi, \delta\bm{u}\right)_{\Delta_l} &= \left\langle\bm{\Phi}\cdot\bm{n}, \widehat{\delta\bm u}\right\rangle_{\partial \Delta_l}, \label{eq: HDG_weak_aux_u}\\
-(\nabla_{\bm{x}}\psi,\bm f^{\delta})_{\Delta_l} + \left\langle\psi,\widehat {\bm f^\delta\cdot\bm{n}}\right\rangle_{\partial \Delta_l} &=- (\psi,\nabla_{\bm{x}}\cdot\bm q^{n+1/2})_{\Delta_l}, \label{eq: HDG_weak_energy}\\
-\left(\nabla_{\bm{x}}\bm{\phi}, \bm{\sigma}^\delta\right)_{\Delta_l}
+\left\langle\bm{\phi}, \widehat{\bm\sigma^\delta\bm{n}}\right\rangle_{\partial \Delta_l}
+(\bm{\phi},\bm{r}^\delta)_{\Delta_l} &=-\left(\bm{\phi}, \nabla_{\bm{x}}\cdot\bm{\Pi}^{n+1/2}\right)_{\Delta_l}-(\bm{\phi},\bm{p}_{\mathrm R}^{n+1/2})_{\Delta_l}, \label{eq: HDG_weak_momentum}
\end{align}
\end{subequations}
where $\widehat{\bm{f}^{\delta}\cdot\bm{n}}$ and
$\widehat{\bm\sigma^{\delta}\bm{n}}$ are the HDG numerical fluxes~\cite{Sevilla2018HDG}. Hybrid traces $\{\widehat{\delta T_{\mathrm N}},\widehat{\delta\bm u}\}$ are determined by imposing conservation of the HDG fluxes across every interior face $\Gamma_c$, i.e., for all $\{\lambda,\bm{\Lambda}\}\in \mathcal{W}_h\times[\mathcal{W}_h]^d$
\begin{subequations}\label{eq: HDG_transmission}
\begin{align}
    \left\langle\lambda, \widehat{\bm{f}^\delta\cdot\bm{n}}^+ + \widehat{\bm{f}^\delta\cdot\bm{n}}^ -\right\rangle_{\Gamma_c} &= 0, \\
    \left\langle\bm{\Lambda}, \widehat{\bm\sigma^\delta\bm n}^+ + \widehat{\bm\sigma^\delta\bm{n}}^-\right\rangle_{\Gamma_c} &= 0,
\end{align}
\end{subequations}
where the superscripts `$\pm$' denote the numerical fluxes evaluated from the elements on both sides of a face. On physical boundaries, the corresponding kinetic boundary data are used. Eq. \eqref{eq: HDG_transmission} is a global problem. Once $\{\widehat{\delta T_{\mathrm N}},\widehat{\delta\bm u}\}$ are solved, the field variables $\{\delta T_{\mathrm{N}},\delta\bm u,\bm g,\bm G\}$ are obtained element-wise from~\eqref{eq: HDG_weak}. That is, the hybrid structure permits static condensation of all element-interior degrees of freedom, thus reduced computational cost compared to the conventional local DG for a second-order system.

\section{Numerical Validation}
\label{sec: numerical_validation}

The suspended monolayer-graphene phonon model, as well as the numerical tests used in~\cite{guo2017heat} are regarded as the benchmark of our numerical validation to (a) examine whether CIS and GSIS approach the same correct solution; and (b) quantify the acceleration produced by GSIS compared to CIS. Additional tests examine unstructured meshes and alternative combinations of isothermal and diffuse adiabatic boundaries. Scale-dependent performance is reported directly in terms of physical lengths $L$/$W$ (longitudinal and transverse sizes of a material), consistent with the interpretation in \Cref{sec: fourier_analysis}.


\subsection{Phonon model and material parameters}
\label{sec: nv_model}

Graphene is a two-dimensional crystal, so its density of states and phase-space integrals are naturally defined per unit area. To report a three-dimensional thermal conductivity, we use the monolayer thickness $h_0=3.35\times10^{-10}\,\mathrm m$ and convert quantities per area to their effective per-volume counterparts by division by $h_0$~\cite{guo2017heat}. This conversion rescales the heat capacity and heat flux, but does not alter the synthetic equation or its closure relations.

Three acoustic branches are retained (see \Cref{sec: fourier_num_evl}), since the optical branches make a negligible contribution to heat transport in the benchmark model~\cite{guo2017heat}. The isotropic $\Gamma$--M dispersions are
\begin{equation}\label{eq: nv_dispersion}
  \omega_{\mathrm{LA}}=c_{\mathrm{LA}}\,k,\qquad
  \omega_{\mathrm{TA}}=c_{\mathrm{TA}}\,k,\qquad
  \omega_{\mathrm{ZA}}=c_{\mathrm{ZA}}\,k^{2}.
\end{equation}
The in-plane branches therefore have constant group speeds $|\bm v|=c_p$, whereas the flexural branch has $|\bm v|_{\mathrm{ZA}}=2\sqrt{c_{\mathrm{ZA}}\omega}$. The corresponding two-dimensional densities of states are $D_{\mathrm{LA/TA}}=k/(2\pi c_p)$ and $D_{\mathrm{ZA}}=1/(4\pi c_{\mathrm{ZA}})$, and the wave number extends to the Brillouin-zone edge $k_m=1.5\times10^{10}\,\mathrm m^{-1}$.

The resistive relaxation rate is the sum of the three-phonon Umklapp and isotope-scattering rates, while the normal process is modeled separately:
\begin{equation}\label{eq: nv_relax}
  \frac1{\tau_{\mathrm U}}=B_{\mathrm U,p}\,\omega^{2}T\,
        e^{-\Theta_p/3T},\quad
  \frac1{\tau_{\mathrm I}}=A_{\mathrm I,p}\,\Gamma S_0\,\omega^{2}D(p,\omega),\quad
  \frac1{\tau_{\mathrm N}}=B_{\mathrm N,p}\,\omega\,T^{3},\quad
  \frac1{\tau_{\mathrm R}}=\frac1{\tau_{\mathrm U}}+\frac1{\tau_{\mathrm I}},
\end{equation}
where $A_{\mathrm I,\mathrm{ZA}}=\pi/2$, $A_{\mathrm I,\mathrm{LA/TA}}=\pi/4$, and the isotope mass variance is $\Gamma=7.541\times10^{-5}$. The remaining coefficients are
\begin{equation}\label{eq: nv_coeffs}
  B_{\mathrm U,p}=\frac{\hbar\gamma_p^{2}}{\overline M\,\Theta_p\,v_{g,p}^{2}},
  \qquad
  B_{\mathrm N,p}=\Bigl(\frac{k_B}{\hbar}\Bigr)^{3}
    \frac{\hbar\gamma_p^{2}\,V_0^{(a_{\mathrm N}+b_{\mathrm N}-2)/3}}
         {\overline M\,v_{g,p}^{a_{\mathrm N}+b_{\mathrm N}}}.
\end{equation}
Here $v_{g,p}$ is the mode group speed from \eqref{eq: nv_dispersion}; it is constant for LA/TA and frequency-dependent for ZA. The parameters in \Cref{tab: nv_params} are used throughout.

\begin{table}
\centering
\renewcommand{\arraystretch}{1.2}
\begin{tabular}{l|c|c|c}
\hline
Parameter & LA & TA & ZA \\
\hline
$c_p$              & $2.13\times10^{4}\,\mathrm{m/s}$ & $1.36\times10^{4}\,\mathrm{m/s}$ & $6.2\times10^{-7}\,\mathrm{m^{2}/s}$\\
$\gamma_p$         & $2$        & $2/3$      & $-1.5$ \\
$\Theta_p$ (K)     & $1826.39$  & $1126.18$  & $623.62$ \\
\hline
\multicolumn{4}{c}{Common parameters}\\
\hline
\multicolumn{2}{l|}{Brillouin-zone edge $k_m$}        & \multicolumn{2}{l}{$1.5\times10^{10}\,\mathrm m^{-1}$}\\
\multicolumn{2}{l|}{Monolayer thickness $h_0$}        & \multicolumn{2}{l}{$3.35\times10^{-10}\,\mathrm m$}\\
\multicolumn{2}{l|}{Per-atom volume $V_0$ / area $S_0$}& \multicolumn{2}{l}{$8.7696\times10^{-30}\,\mathrm m^{3}$ / $2.62\times10^{-20}\,\mathrm m^{2}$}\\
\multicolumn{2}{l|}{Mean atomic mass $\overline M$}   & \multicolumn{2}{l}{$1.9945\times10^{-26}\,\mathrm{kg}$}\\
\multicolumn{2}{l|}{N-scattering exponents $(a_{\mathrm N},b_{\mathrm N})$} & \multicolumn{2}{l}{$(1,3)$}\\
\hline
\end{tabular}
\caption{Phonon-model parameters for suspended monolayer graphene from~\cite{guo2017heat}.}
\label{tab: nv_params}
\end{table}

\subsection{Numerical settings and validation cases}
\label{sec: nv_settings}

\paragraph{Quadrature.}
All integrals with respect to the phonon spectrum and propagation direction are evaluated using the numerical quadrature. In the benchmark cases, each branch uses an eight-node Gauss--Legendre rule on $[0,\omega_{\max}^p]$, matching the benchmark spectral resolution, while the polar angle is represented by $M_\Omega$ ordinates as listed in \Cref{tab: nv_cases} and the mid-point rule is used for the corresponding integrals. Note that several $\tau_{\mathrm N}$-weighted ZA moments are sensitive to low-frequency quadrature points because $(k/\omega)^2\propto\omega^{-1}$ and the ZA group speed makes $\tau_{\mathrm N}^{-1}\propto\omega^{-1}$ as $\omega\to0$. The benchmark-consistent eight-node rule is therefore retained for every benchmark calculation.


\paragraph{Spatial resolution.}
The cases in benchmark validation suite are discretized by structured triangular meshes formed by diagonally bisecting each cell of an underlying rectangular partition. Cases B-1, B-2, and D use uniform partitions with $20\times20$. Case C has the same resolution, with uniform spacing in the periodic $x$ direction and refinement toward the two adiabatic boundaries in the $y$ direction. The extra cases are tested on the unstructured triangular meshes. CIS and GSIS use identical meshes. The degree of polynomials for spatial approximation is $K=3$, and the HDG global matrix is assembled and factorized once by a sparse direct solver.

\paragraph{Iteration stopping rule and transport sweeping.}
Both GSIS and CIS use the same stopping criterion~\eqref{eq: stop_criterion} choosing $\mathrm{tol}=10^{-7}$. Otherwise, it ends in the iteration cap $N_{\max}=2\times10^{5}$ and is determined to be non-converged. Thus, every run stops either after satisfying the prescribed tolerance or after reaching $N_{\max}$. When solving the kinetic equation, the transport term is treated implicitly. To avoid the assembly of a global matrix, the transport sweep technique is used by marching along the phonon propagation direction from a given boundary condition. 


\paragraph{Timing protocol.}
All timings were obtained from Release builds on a 13th-generation Intel Core i9-13900K workstation with 24 cores at $3.00\,\mathrm{GHz}$. At most three tests were run concurrently, each using 8 OpenMP threads, and the BLAS libraries were restricted to one thread. CIS and GSIS used identical discretizations. Iteration count is therefore used as the primary algorithmic measure, while wall-clock time is reported as an implementation- and hardware-dependent measure.

\paragraph{Validation cases.}
We consider five different cases, marked `A', `B-1', `B-2', `C' and `D' for benchmark testing. The geometries, boundary conditions, parameter ranges, and angular resolutions are summarized in \Cref{tab: nv_cases}; the phonon spectrum and angular resolutions follow the settings in~\cite{guo2017heat}.

\begin{table}
\centering
\renewcommand{\arraystretch}{1.25}
\begin{tabular}{c|p{9cm}|p{4cm}|c}
\hline
Case & Geometry / boundary conditions & Parameter sweep & $M_{\Omega}$ \\
\hline
A   & Bulk limit (analytic Callaway; no PDE solve)
    & $T\in[100,800]~\mathrm K$
    & --   \\
B-1 & 2D rectangle; isothermal ends (left/right), periodic transverse boundaries (top/bottom)
    & $L\in[10^{-3},10^{2}]~\mu\mathrm m$
    & $32$ \\
B-2 & 2D rectangle; isothermal ends (left/right), periodic transverse boundaries (top/bottom)
    & $T\in[300,600]~\mathrm K$, \newline $L\in\{3,5,10\}~\mu\mathrm m$
    & $32$ \\
C   & 2D rectangle; periodic heat-flux boundaries (left/right), diffuse adiabatic walls (top/bottom)
    & $W\in[10^{-2},10^{2}]~\mu\mathrm m$
    & $96$ \\
D   & 2D rectangle; isothermal ends (left/right), diffuse adiabatic walls (top/bottom)
    & $L\in[10^{-2},50]~\mu\mathrm m$
    & $48$ \\
\hline
\end{tabular}
\caption{Benchmark cases and angular resolutions. All cases use 8 Gauss--Legendre frequency nodes per branch. Cases B-1, B-2, C, and D use $K=3$ nodal DG, $\Delta T=1~\mathrm K$, and the spatial resolution $20\times20$, corresponding to 800 triangular elements after diagonal bisection of the rectangular cells. Cases B-1, B-2, and D use uniform partitions, whereas Case C is refined toward its adiabatic walls.}
\label{tab: nv_cases}
\end{table}

\paragraph{Effective thermal conductivity.}
For the finite-domain cases, the diagnostic quantity is
\begin{equation}\label{eq: nv_kappa_eff}
  \kappa_{\mathrm{eff}}=\frac{\bar{q}L}{h_0\Delta T},
  \qquad
  \bar{q}=\frac{1}{|\Delta|}\int_\Delta q\,\mathrm d\bm x,
\end{equation}
where $q$ is the converged heat-flux component in the transport direction. In GSIS it is taken from $\bm M^{n+1}$ after the HDG solve and macroscopic update; in CIS it is the direct kinetic moment. Division by $h_0$ converts the two-dimensional flux to an effective three-dimensional value in $\mathrm{W\,m^{-1}K^{-1}}$. For homogeneous Case~A, the conductivity is evaluated directly from the analytic Callaway decomposition $\kappa=\kappa_{\mathrm{SMRT}}+\kappa_{\mathrm C}$~\cite{guo2017heat} on the same spectral grid, without a spatial solve:
\begin{equation}
    \kappa_{\mathrm{SMRT}} = \frac{1}{2h_0} \left\langle \tau_{\mathrm{C}} |\bm v|^2 C_{p,\omega} \right\rangle_{p}, \qquad \kappa_{\mathrm C} = \frac{1}{2h_0}\frac{\left\langle \frac{\tau_{\mathrm{C}}}{\tau_{\mathrm{N}}} \frac{|\bm k||\bm v|}{\omega} C_{p,\omega} \right\rangle_{p}}{\left\langle \frac{\tau_{\mathrm{C}}}{\tau_{\mathrm{N}}\tau_{\mathrm{R}}} \frac{|\bm k|^2}{\omega^2} C_{p,\omega} \right\rangle_{p}}
\end{equation}

\subsection{Benchmark results}
\label{sec: nv_results}

\paragraph{Bulk thermal conductivity}
\label{sec: nv_A}

\begin{figure}
\centering
\includegraphics[width=0.6\textwidth]{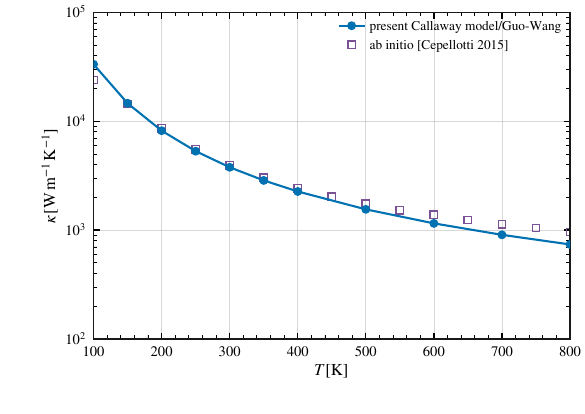}
\caption{Case~A: bulk $\kappa(T)$ from the present Callaway model compared with the Callaway-model and \emph{ab initio} reference data reported in~\cite{guo2017heat}. This case is analytic and independent of the iterative solver.}
\label{fig: nv_A}
\end{figure}

\Cref{fig: nv_A} compares the bulk conductivity obtained on the present spectral grid with the reference data digitized from~\cite{guo2017heat}. The calculation reproduces the reference Callaway-model values. Its departure from the \emph{ab initio} data at high temperature is consistent with the difference reported for the Callaway approximation in the benchmark. Because no spatial or iterative solve is involved, this test isolates the material model and frequency quadrature from the numerical solvers considered below.

\paragraph{Finite-domain validation and solver consistency}
\label{sec: nv_consistency}

\begin{figure}
\centering
\subfloat[B-1: $\kappa_{\mathrm{eff}}(L)$ at $T=300~\mathrm K$.]{
  \includegraphics[width=0.48\textwidth]{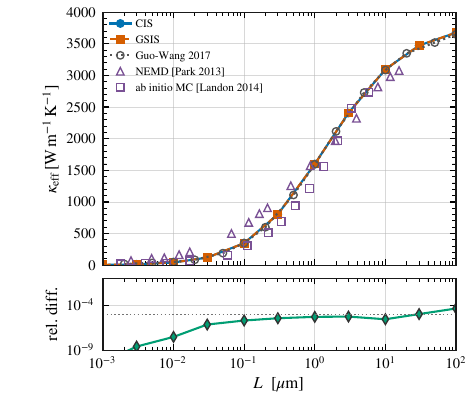}}
\hfill
\subfloat[B-2: $\kappa_{\mathrm{eff}}(T)$ at $L=3,5,10~\mu\mathrm m$.]{
  \includegraphics[width=0.48\textwidth]{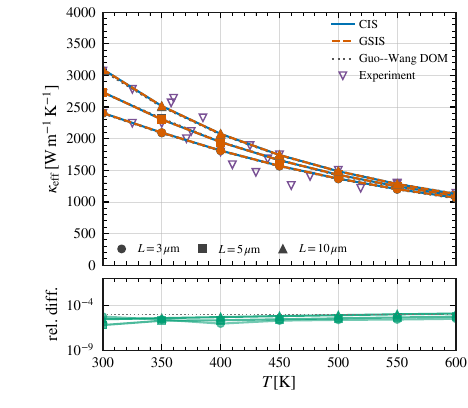}}\\[2pt]
\subfloat[C: Poiseuille $\kappa_{\mathrm{eff}}(W)$ at $T=300~\mathrm K$.]{
  \includegraphics[width=0.48\textwidth]{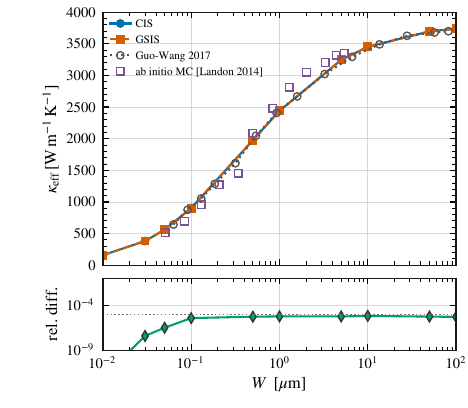}}
\hfill
\subfloat[D: 2D rectangle $\kappa_{\mathrm{eff}}(L)$ at $W=1.5~\mu\mathrm m$.]{
  \includegraphics[width=0.48\textwidth]{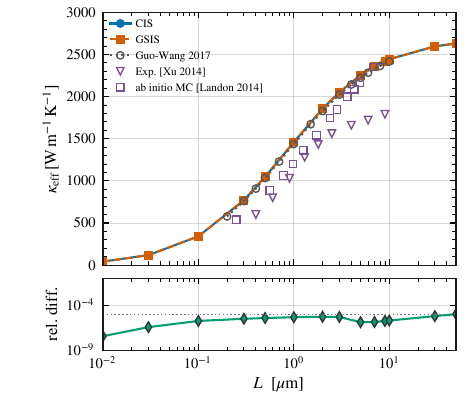}}
\caption{Effective thermal conductivity for the four finite-domain validation cases. CIS and GSIS are compared with data digitized from~\cite{guo2017heat}, including its DOM solution and the reported \emph{ab initio} Monte Carlo, NEMD, and experimental results. The lower strip in each panel shows the pointwise CIS--GSIS relative difference at jointly converged points.}
\label{fig: nv_kappa_curves}
\end{figure}

\Cref{fig: nv_kappa_curves} covers the finite-ribbon length sweep at $300~\mathrm K$ (B-1), the temperature sweep at three fixed lengths (B-2), the Poiseuille width sweep (C), and the two-dimensional length sweep with diffuse lateral walls (D). The calculations span strongly size-dependent transport, collective Poiseuille behavior, and the approach toward bulk conduction without assigning a single transport regime to the full non-gray spectrum. At all $56$ jointly converged points, the CIS and GSIS curves overlap at the plotting scale and follow the reference DOM results closely.

The relative conductivity difference between the two solvers has a median of $3.38\times10^{-6}$ and a maximum of $4.39\times10^{-5}$ at the prescribed tolerance. This small discrepancy is a finite-iteration effect. In GSIS, the stopping rule is applied to the fully updated macroscopic state $\bm M^{n+1}$, while the kinetic distribution is advanced as $e^{n+1}=e^{n+1/2}$; before the residual vanishes, the updated macroscopic fields need not coincide exactly with the direct moments of the kinetic intermediate-step. Tightening the tolerance reduces this discrepancy. At the fixed point, the kinetic residual and the synthetic increment vanish, so CIS and GSIS recover the same discrete solution.

\paragraph{Iterative acceleration}
\label{sec: nv_acceleration}

\begin{figure}
\centering
\subfloat[B-1: iterations vs.\ $L$ ($T=300~\mathrm K$).]{
  \includegraphics[width=0.48\textwidth]{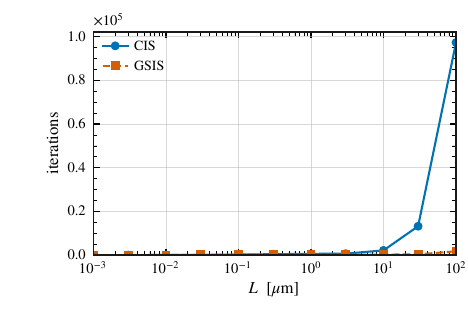}}
\hfill
\subfloat[B-2: iterations vs.\ $T$ ($L=3,5,10~\mu\mathrm m$).]{
  \includegraphics[width=0.48\textwidth]{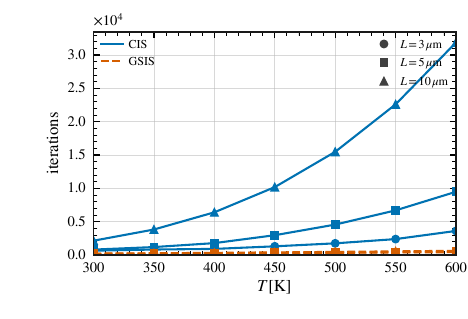}}\\[2pt]
\subfloat[C: iterations vs.\ $W$ ($T=300~\mathrm K$).]{
  \includegraphics[width=0.48\textwidth]{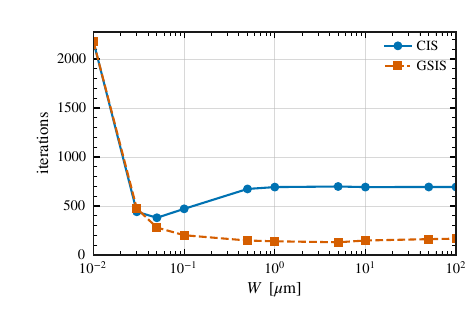}}
\hfill
\subfloat[D: iterations vs.\ $L$ ($W=1.5~\mu\mathrm m$).]{
  \includegraphics[width=0.48\textwidth]{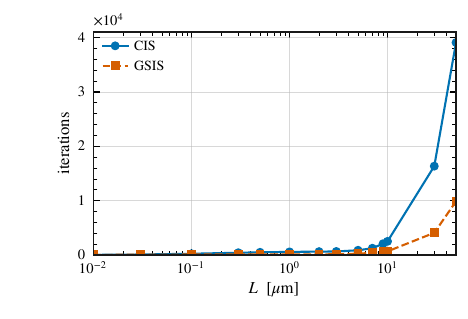}}
\caption{Outer-iteration counts of CIS and GSIS at the common tolerance $\mathrm{tol}=10^{-7}$ for the four validation cases. The reduction produced by GSIS becomes most pronounced at parameter points where CIS develops slow long-wavelength error. All displayed benchmark points converge before reaching $N_{\max}$.}
\label{fig: nv_iters}
\end{figure}

\begin{figure}
\centering
\subfloat[B-1: wall time vs.\ $L$ ($T=300~\mathrm K$).]{
  \includegraphics[width=0.46\textwidth]{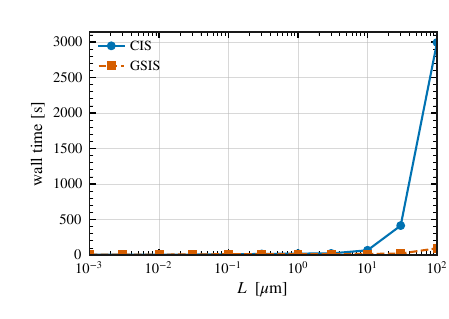}}
\hfill
\subfloat[B-2: wall time vs.\ $T$ ($L=3,5,10~\mu\mathrm m$).]{
  \includegraphics[width=0.46\textwidth]{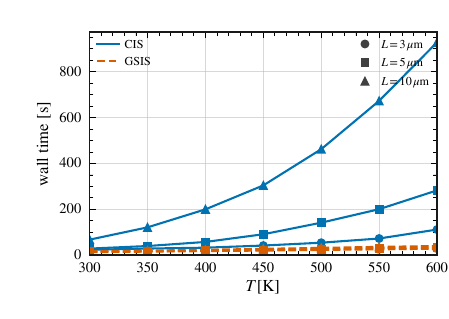}}\\[2pt]
\subfloat[C: wall time vs.\ $W$ ($T=300~\mathrm K$).]{
  \includegraphics[width=0.46\textwidth]{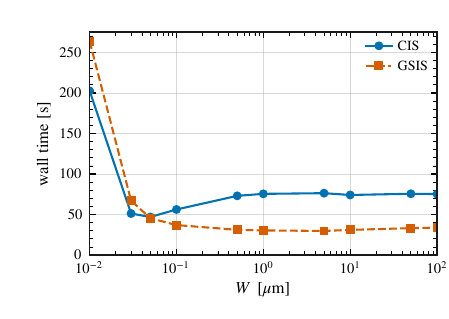}}
\hfill
\subfloat[D: wall time vs.\ $L$ ($W=1.5~\mu\mathrm m$).]{
  \includegraphics[width=0.46\textwidth]{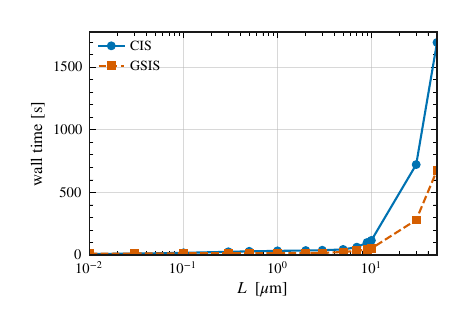}}
\caption{Wall-clock solution times for the four validation cases under identical meshes and process/thread allocations. Each GSIS iteration includes an additional HDG macroscopic solve, but the reduced iteration count yields a net time saving at most of the more slowly converging CIS points.}
\label{fig: nv_walltime}
\end{figure}

The iteration counts in \Cref{fig: nv_iters} provide the implementation-independent comparison. In Cases B-1 and D, CIS and GSIS require comparable numbers of iterations at the smallest $L$, while the separation grows as $L$ increases; in Case B-2, the acceleration likewise becomes more pronounced at the more slowly converging combinations of $T$ and $L$. Case C exhibits a different dependence on the transverse width: the iteration counts are relatively large at the smallest $W$, decrease as $W$ increases, and then approach a nearly width-independent level over the larger-$W$ range. We therefore treat this width dependence as a distinct feature of Case C rather than imposing the monotonic length-scale trend observed in the other sweeps. The additional HDG solve makes each GSIS iteration more expensive, so the wall-time gain in \Cref{fig: nv_walltime} is generally smaller than the iteration-count gain and can be negative at the easiest points. Every benchmark point used in the figures and statistics satisfies \eqref{eq: stop_criterion} before reaching $N_{\max}$.

The largest iteration reduction occurs in Case B-1 at $L=100~\mu\mathrm m$: CIS requires $97\,346$ iterations and $4196.3~\mathrm s$, whereas GSIS converges in $1755$ iterations and $118.3~\mathrm s$, corresponding to speed-ups of $55.5\times$ and $35.5\times$, respectively. The largest wall-time gain occurs in Case B-2 at $L=10~\mu\mathrm m$ and $T=600~\mathrm K$, where the iteration count decreases from $31\,885$ to $649$ and the time from $2125.0$ to $57.0~\mathrm s$, giving speed-ups of $49.1\times$ and $37.3\times$. Across all $56$ jointly converged points, the median iteration-count and wall-time speed-ups are $4.26\times$ and $2.46\times$, their maxima are $55.5\times$ and $37.3\times$, and the ratio of the summed CIS and GSIS wall times is $5.54\times$. The wall-time ratios remain specific to the hardware and implementation, whereas the iteration-count ratios more directly characterize the algorithmic acceleration.

\subsection{Tests on unstructured meshes and mixed boundary conditions}
\label{sec: nv_examples}

The benchmark calculations use axis-aligned rectangular domains discretized by structured triangular meshes formed by diagonal bisection of an underlying rectangular partition. To examine the implementation on general meshes and with additional boundary-condition combinations, we consider two configurations on a unit square at $T_0=300~\mathrm K$, with $T_h=300.5~\mathrm K$ and $T_c=299.5~\mathrm K$, discretized by unstructured triangular meshes:
\begin{enumerate}
    \item \texttt{iso4}: all four walls are isothermal;
    \item \texttt{cavity}: the left and right walls are isothermal, while the top and bottom walls and a centered circular hole of radius $0.2~\mu\mathrm m$ are diffuse non-thermalizing adiabatic boundaries.
\end{enumerate}
Both configurations use 20 frequency nodes per branch, 48 angular ordinates, and the stopping rule \eqref{eq: stop_criterion} with the same $N_{\max}$. The degree of polynomials in DG approximation is $K=4$.

\paragraph{Unstructured-mesh fields.}

\begin{figure}
\centering
\subfloat[\texttt{iso4}: CIS (left) and GSIS (right).]{
  \includegraphics[width=0.48\textwidth]{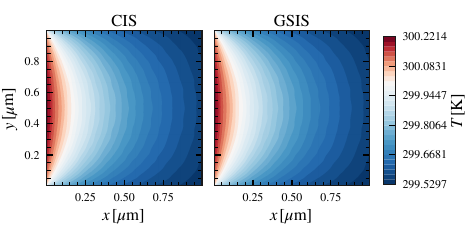}}
\hfill
\subfloat[\texttt{cavity}: CIS (left) and GSIS (right).]{
  \includegraphics[width=0.48\textwidth]{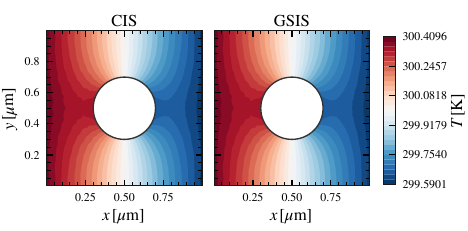}}
\caption{Temperature fields on unstructured meshes for the $1~\mu\mathrm m$ \texttt{iso4} and \texttt{cavity} configurations. The CIS and GSIS fields are visually indistinguishable at the plotted scale.}
\label{fig: nv_ex_cntr}
\end{figure}

The paired fields in \Cref{fig: nv_ex_cntr} show that the two iterative schemes produce the same resolved temperature distribution on the plotted scale. The cavity configuration also exercises the diffuse non-thermalizing boundary treatment on an internal curved boundary.

\paragraph{Scale sweep for the \texttt{iso4} configuration.}

\begin{figure}
\centering
\subfloat[Effective conductivity $\kappa_{\mathrm{eff}}(L)$.]{
  \includegraphics[width=0.32\textwidth]{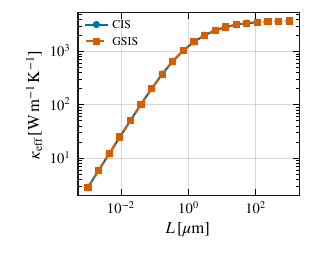}}
\hfill
\subfloat[Outer-iteration counts.]{
  \includegraphics[width=0.32\textwidth]{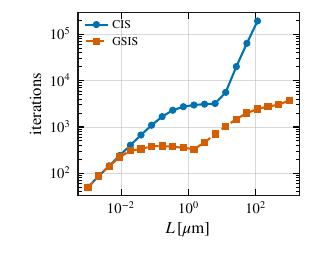}}
\hfill
\subfloat[CIS-to-GSIS iteration-count ratio.]{
  \includegraphics[width=0.32\textwidth]{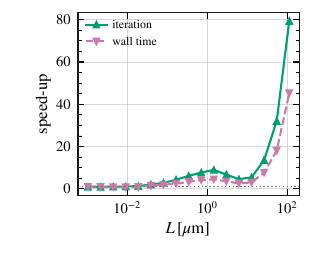}}
\caption{Size effect and convergence for the \texttt{iso4} configuration over $L=1~\mathrm{nm}$ to $1~\mathrm{mm}$. Ratios involving a CIS run terminated at $N_{\max}$ are lower bounds rather than speed-ups between two converged runs.}
\label{fig: nv_ex_scale}
\end{figure}

For the \texttt{iso4} configuration at $K=4$, $20$ logarithmically spaced lengths from $1~\mathrm{nm}$ to $1~\mathrm{mm}$ are used to examine the size dependence (\Cref{fig: nv_ex_scale}). The effective conductivity increases from the strongly size-dependent small-$L$ response toward the bulk plateau, reaching approximately $3.7\times10^{3}~\mathrm{W\,m^{-1}K^{-1}}$ at $L=1~\mathrm{mm}$ and approaching the analytic Case~A value. Wherever both solvers converge, they give the same $\kappa_{\mathrm{eff}}$ at the reported precision.

The same sweep provides an additional comparison of iterative performance. Over the jointly converged range through $L=10^{2}~\mu\mathrm m$, the CIS count grows rapidly with the characteristic length, as anticipated from the long-wavelength Fourier analysis, whereas the GSIS count remains of order $10^{3}$. At the first sampled length above $10^{2}~\mu\mathrm m$, $L\approx113~\mu\mathrm m$, CIS reaches $N_{\max}=2\times10^{5}$ without satisfying \eqref{eq: stop_criterion}, while GSIS converges in approximately $2.4\times10^{3}$ iterations. The ratio of the CIS cap to the converged GSIS count already exceeds $80$, but it is only a lower bound because the CIS run is nonconverged. At every larger sampled length through $1~\mathrm{mm}$, GSIS continues to converge within the common cap, whereas CIS does not. The same acceleration trend therefore persists throughout the sampled sizes for this boundary configuration.

\subsection{Discussion}
\label{sec: nv_summary}

The numerical tests collectively verify the main objectives of the proposed formulation and implementation. The bulk calculation in Case A reproduces the Callaway-model conductivity obtained with the adopted spectral representation, while the finite-domain Cases B--D recover the corresponding benchmark trends over their respective length, width, and temperature sweeps. At all jointly converged points, CIS and GSIS give essentially identical effective conductivities, consistent with their convergence toward the same discrete kinetic solution. The additional \texttt{iso4} and \texttt{cavity} tests extend this agreement to unstructured triangular meshes and different combinations of isothermal and diffuse adiabatic boundaries, including an internal curved boundary. Moreover, the \texttt{iso4} scale sweep connects the strongly size-dependent response at small \(L\) to the bulk-conductivity plateau at large \(L\), providing a further consistency check over a substantially wider range of physical scales.

The convergence results confirm the intended acceleration provided by the macroscopic synthetic equations. The benefit is modest when CIS already converges rapidly, but becomes substantial at the parameter points where the kinetic iteration requires many sweeps: among the benchmark cases, the maximum reductions in iteration count and wall time are \(55.5\times\) and \(37.3\times\), respectively. The extended \texttt{iso4} sweep makes this contrast more pronounced: CIS reaches the common cap \(N_{\max}=2\times10^{5}\) from \(L\approx113~\mu\mathrm m\) onward, whereas GSIS continues to converge with iteration counts of order \(10^{3}\) through \(L=1~\mathrm{mm}\). Since each GSIS iteration includes an additional HDG solve, its wall-time gain is naturally smaller than its iteration-count reduction and can disappear at the easiest parameter points. Case C also shows that the iteration count need not vary monotonically with a single geometric length: convergence is slower at the smallest \(W\), while the counts decrease and then become nearly insensitive to further increases in \(W\). We regard this as a case-dependent feature of the present geometry and boundary conditions; it does not alter the overall observation that GSIS preserves the converged solution while substantially reducing the iterative cost where CIS becomes expensive.

\section{Conclusion and Outlook}
\label{sec: conclusion_outlook}

In this work, we have developed a general synthetic iterative scheme for the stationary, linearized non-gray Callaway phonon Boltzmann transport equation with mode-dependent normal and resistive relaxation times. The proposed GSIS framework couples the kinetic equation with the macroscopic synthetic equations derived from it. GSIS efficiently corrects the slowly converging macroscopic error modes while retaining the mode-resolved kinetic description and its non-equilibrium information. Incorporated with the DG and HDG spatial discretization, the present work therefore provides a high-order computational framework for the mode-dependent phonon transport across multiscale regimes.

The synthetic equations were constructed from the exact energy and quasi-momentum conservation, with first-order constitutive relations obtained from a Chapman--Enskog analysis and higher-order non-equilibrium terms evaluated directly from the kinetic solution. The resulting iteration was implemented by combining a nodal DG discretization of the kinetic equation with an HDG discretization of the synthetic macroscopic system. A branch- and frequency-resolved Fourier analysis identified the long-wavelength mode responsible for the deterioration of CIS at large characteristic lengths and showed that the macroscopic correction prevents the corresponding GSIS contraction factor from approaching unity. The asymptotic analysis further established fixed-point compatibility between the kinetic and synthetic fields and showed that, under the corresponding scalings, the converged formulation recovers a Guyer--Krumhansl-like hydrodynamic equation and Fourier's heat-conduction equation in the hydrodynamic and diffusive limits, respectively. Numerical experiments with the non-gray graphene model reproduced established benchmark results and confirmed that CIS and GSIS approach the same correct discrete kinetic solution. Across the 56 jointly converged benchmark points, GSIS achieved maximum reductions of $55.5\times$ in iteration count and $37.3\times$ in wall-clock time. Tests on unstructured meshes and mixed boundary conditions further demonstrated the flexibility of the finite-element implementation, while the extended $1~\mathrm{nm}$--$1~\mathrm{mm}$ size sweep showed that GSIS continues to converge with iteration counts of order $10^3$ even when CIS reaches the prescribed $2\times10^5$ iteration cap.

The present work also provides a basis for future work in related area. Natural extensions include three-dimensional geometries and anisotropic phonon dispersions, as well as transient and nonlinear formulations for problems with larger temperature variations and temperature-dependent material properties.

\appendix
\section{Transport coefficients} \label{sec: A1}
Transport coefficients in Eqs.~\eqref{eq: CE_0th_macro} and~\eqref{eq: CE_CR} are
\begin{equation}\label{eq: C_Constants}
    C_{vk} = \left\langle \frac{|\bm{v}|\,|\bm{k}|\,C_{p,\omega}}{\omega}\right\rangle_p, \quad C_k = \left\langle \frac{|\bm{k}|^2\,C_{p,\omega}}{\omega^2}\right\rangle_p,\quad
    C_{\tau_{\mathrm{R}}}^{1} = \left\langle \frac{|\bm{k}|^2}{\omega^2}\frac{C_{p,\omega}}{\tau_{\mathrm{R}}}\right\rangle_p,
\end{equation}

\begin{equation}\label{eq: kappa_N}
    \kappa_{\mathrm{N}} = \frac{1}{d}\left[\left\langle \tau_{\mathrm{N}}|\bm{v}|^2 C_{p,\omega}\right\rangle_p - \frac{C_{vk}}{C_k}\left\langle \tau_{\mathrm{N}}\frac{|\bm{v}|\,|\bm{k}|\,C_{p,\omega}}{\omega}\right\rangle_p\right],
\end{equation}
\begin{equation}\label{eq: mu_N_1_2}
    \mu_{\mathrm{N}}^{1} = \frac{T_0}{d(d+2)}\left\langle \tau_{\mathrm{N}}|\bm{v}|^2\frac{|\bm{k}|^2 C_{p,\omega}}{\omega^2}\right\rangle_p,\qquad \mu_{\mathrm{N}}^{2} = \frac{T_0}{d^2}\frac{C_{vk}}{C_V}\left\langle \tau_{\mathrm{N}}\frac{|\bm{v}|\,|\bm{k}|\,C_{p,\omega}}{\omega}\right\rangle_p,
\end{equation}
\begin{equation}\label{eq: eta}
    \eta = \frac{1}{d}\left[\left\langle \frac{\tau_{\mathrm{N}}}{\tau_{\mathrm{R}}}\frac{|\bm{v}|\,|\bm{k}|\,C_{p,\omega}}{\omega}\right\rangle_p -\left\langle \frac{C_{vk}}{C_k}\sum_p \int \frac{\tau_{\mathrm{N}}}{\tau_{\mathrm{R}}}\frac{|\bm{k}|^2 C_{p,\omega}}{\omega^2}\right\rangle_p\right],
\end{equation}
\begin{equation}\label{eq: gamma}
    \gamma = \frac{T_0}{d C_{\tau_{\mathrm{R}}}}\left[\left\langle\frac{C_{vk}}{C_V}\sum_p \int \frac{\tau_{\mathrm{N}}}{\tau_{\mathrm{R}}}C_{p,\omega}\right\rangle_p - \left\langle \frac{\tau_{\mathrm{N}}}{\tau_{\mathrm{R}}}\frac{|\bm{v}|\,|\bm{k}|\,C_{p,\omega}}{\omega}\right\rangle_p\right].
\end{equation}

\section{Calculations in Fourier's analysis} \label{appendix: Fourier_analysis_cal}

\subsection{Analytical integrals}
\label{appendix: analytical_integrals}

\paragraph{Closed-form angular resolvents}
For each phonon mode $(p,\omega)$, define the dimensionless Fourier-scale parameter
\begin{equation}\label{eq: beta_phi_def}
    \beta_{p,\omega}
    :=\xi\ell_{p,\omega}
    =\frac{2\pi\ell_{p,\omega}}{L_\xi},
    \qquad
    s_{p,\omega}:=\sqrt{1+\beta_{p,\omega}^2},
    \qquad
    \Phi_{p,\omega}:=1-\frac{1}{s_{p,\omega}}.
\end{equation}
Then
\begin{equation}\label{eq: Ij}
    \mathcal I_{p,\omega}^{0} = \frac{\tau_{\mathrm C}}{s_{p,\omega}}, \qquad
    \mathcal I_{p,\omega}^{1} = \frac{\Phi_{p,\omega}}{\mathrm i\xi v_{p,\omega}} = -\mathrm i\frac{\Phi_{p,\omega}}{\xi v_{p,\omega}}, \qquad
    \mathcal I_{p,\omega}^{2} = \frac{\tau_{\mathrm C}}{s_{p,\omega}(s_{p,\omega}+1)}, \qquad
    \mathcal J_{p,\omega} = \frac{\tau_{\mathrm C}}{s_{p,\omega}+1}.
\end{equation}
In particular, as $\beta_{p,\omega}\to0$,
\begin{equation}\label{eq: Ij_zero}
    \mathcal I_{p,\omega}^{0}\to\tau_{\mathrm C},
    \qquad
    \mathcal I_{p,\omega}^{1}\to0,
    \qquad
    \mathcal I_{p,\omega}^{2}\to\frac{\tau_{\mathrm C}}{2},
    \qquad
    \mathcal J_{p,\omega}\to\frac{\tau_{\mathrm C}}{2}.
\end{equation}

\begin{proof}
For a fixed mode $(p,\omega)$, set $P:=\tau_{\mathrm C}^{-1}>0$, $Q:=\mathrm i\xi v_{p,\omega}$. The denominator in \eqref{eq: resolvent_def} is $P+Q\cos\theta$. The standard full-period integral, continued to the present purely imaginary $Q$, gives
\begin{equation}
    \frac{1}{2\pi}\int_0^{2\pi}
    \frac{\dd\theta}{P+Q\cos\theta}
    =\frac{1}{\sqrt{P^2-Q^2}},
\end{equation}
where the positive real square root is selected. Since $P^2-Q^2 = \tau_{\mathrm C}^{-2}+\xi^2v_{p,\omega}^2 = \tau_{\mathrm C}^{-2}(1+\beta_{p,\omega}^2)$, this yields the first identity in \eqref{eq: Ij}. For the first and second integrals,
\begin{equation}
    \frac{\cos\theta}{P+Q\cos\theta} = \frac{1}{Q}\left(1-\frac{P}{P+Q\cos\theta}\right) \qquad \text{ and } \qquad
    \frac{\cos^2\theta}{P+Q\cos\theta} = \frac{\cos\theta}{Q} -\frac{P}{Q}\frac{\cos\theta}{P+Q\cos\theta},
\end{equation}
yield the second and third identities in \eqref{eq: Ij}. 
\begin{equation}
    \mathcal I_{p,\omega}^{1} = \frac{1}{Q}\left(1-\frac{P}{\sqrt{P^2-Q^2}}\right) = \frac{1-s_{p,\omega}^{-1}}{\mathrm i\xi v_{p,\omega}} = \frac{\Phi_{p,\omega}}
    {\mathrm i\xi v_{p,\omega}}, \qquad
    \mathcal I_{p,\omega}^{2} = -\frac{P}{Q}\mathcal I_{p,\omega}^{1} = \frac{\Phi_{p,\omega}}{\xi^2v_{p,\omega}^2\tau_{\mathrm C}} =\frac{\tau_{\mathrm{C}}}{s_{p,\omega}(s_{p,\omega}+1)}
\end{equation}
where the last equality holds by $\beta_{p,\omega}=\xi v_{p,\omega}\tau_{\mathrm C}$ and $\Phi_{p,\omega}/\beta_{p,\omega}^2 = (1-s_{p,\omega}^{-1})/(s_{p,\omega}^2-1) = 1/s_{p,\omega}(s_{p,\omega}+1)$. Finally,
\begin{equation}
    \mathcal J_{p,\omega} = \mathcal I_{p,\omega}^{0}-\mathcal I_{p,\omega}^{2} = \frac{\tau_{\mathrm C}}{s_{p,\omega}+1},
\end{equation}
which proves the last identity in \eqref{eq: Ij}. Taking $\beta_{p,\omega}\to0$ gives \eqref{eq: Ij_zero}.
\end{proof}

\subsection{Entries of the contraction matrix}
\label{appendix: contraction_matrix_entries}

The entries of the contraction matrix $\bm G_{\mathrm{CIS}}$ [see, Eq.~\eqref{eq: G_CIS}] are
\begin{equation}\label{eq: Gij}
\begin{alignedat}{3}
    G_{11}&=\frac{1}{C_{\tau_{\mathrm R}}}\left\langle\frac{C_{p,\omega}\mathcal I^0_{p,\omega}}{\tau_{\mathrm R}^2}\right\rangle_{p},
    &\quad
    G_{12}&=\frac{1}{C_{\tau_{\mathrm R}}}\left\langle\frac{C_{p,\omega}\mathcal I^0_{p,\omega}}{\tau_{\mathrm R}\tau_{\mathrm N}}\right\rangle_{p},
    &\quad
    G_{13}&=\frac{T_0}{C_{\tau_{\mathrm R}}}\left\langle\frac{k_{p,\omega}C_{p,\omega}\mathcal I^1_{p,\omega}}
    {\omega\tau_{\mathrm R}\tau_{\mathrm N}}\right\rangle_{p},\\[4pt]
    G_{21}&=\frac{1}{C_{\tau_{\mathrm N}}}\left\langle\frac{C_{p,\omega}\mathcal I^0_{p,\omega}}{\tau_{\mathrm R}\tau_{\mathrm N}}\right\rangle_{p},
    &\quad
    G_{22}&=\frac{1}{C_{\tau_{\mathrm N}}}\left\langle\frac{C_{p,\omega}\mathcal I^0_{p,\omega}}{\tau_{\mathrm N}^2}\right\rangle_{p},
    &\quad
    G_{23}&=\frac{T_0}{C_{\tau_{\mathrm N}}}\left\langle\frac{k_{p,\omega}C_{p,\omega}\mathcal I^1_{p,\omega}}
    {\omega\tau_{\mathrm N}^2}\right\rangle_{p},\\[4pt]
    G_{31}&=\frac{d}{T_0C_{\tau_{\mathrm N}}^1}\left\langle\frac{k_{p,\omega}C_{p,\omega}\mathcal I^1_{p,\omega}}
    {\omega\tau_{\mathrm R}\tau_{\mathrm N}}\right\rangle_{p},
    &
    G_{32}&=\frac{d}{T_0C_{\tau_{\mathrm N}}^1}\left\langle\frac{k_{p,\omega}C_{p,\omega}\mathcal I^1_{p,\omega}}
    {\omega\tau_{\mathrm N}^2}\right\rangle_{p},
    &
    G_{33}&=\frac{d}{C_{\tau_{\mathrm N}}^1}\left\langle\frac{k_{p,\omega}^2C_{p,\omega}\mathcal I^2_{p,\omega}}
    {\omega^2\tau_{\mathrm N}^2}\right\rangle_{p},\\[4pt]
    G_{44}&=\frac{d}{C_{\tau_{\mathrm N}}^1}\left\langle\frac{k_{p,\omega}^2C_{p,\omega}\mathcal J_{p,\omega}}
    {\omega^2\tau_{\mathrm N}^2}\right\rangle_{p}.
    &&&&
\end{alignedat}
\end{equation}


\bibliographystyle{elsarticle-num}
\bibliography{Ref}

@article{chen2021nonfourier,
  author  = {Chen, Gang},
  title   = {Non-Fourier Phonon Heat Conduction at the Microscale and Nanoscale},
  journal = {Nature Reviews Physics},
  volume  = {3},
  pages   = {555--569},
  year    = {2021},
  doi     = {10.1038/s42254-021-00334-1}
}

@book{chen2005nanoscale,
  title={Nanoscale energy transport and conversion: a parallel treatment of electrons, molecules, phonons, and photons},
  author={Chen, Gang},
  year={2005},
  publisher={Oxford university press},
  doi= {10.1093/oso/9780195159424.001.0001}
}

@article{guo2015hydrodynamics,
  author  = {Guo, Yangyu and Wang, Moran},
  title   = {Phonon Hydrodynamics and Its Applications in Nanoscale Heat Transport},
  journal = {Physics Reports},
  volume  = {595},
  pages   = {1--44},
  year    = {2015},
  doi     = {10.1016/j.physrep.2015.07.003}
}

@article{callaway1959model,
  author  = {Callaway, Joseph},
  title   = {Model for Lattice Thermal Conductivity at Low Temperatures},
  journal = {Physical Review},
  volume  = {113},
  number  = {4},
  pages   = {1046--1051},
  year    = {1959},
  doi     = {10.1103/PhysRev.113.1046}
}

@article{guyer1966solution,
  author  = {Guyer, R. A. and Krumhansl, J. A.},
  title   = {Solution of the Linearized Phonon Boltzmann Equation},
  journal = {Physical Review},
  volume  = {148},
  number  = {2},
  pages   = {766--778},
  year    = {1966},
  doi     = {10.1103/PhysRev.148.766}
}

@article{guyer1966thermal,
  author  = {Guyer, R. A. and Krumhansl, J. A.},
  title   = {Thermal Conductivity, Second Sound, and Phonon Hydrodynamic Phenomena in Nonmetallic Crystals},
  journal = {Physical Review},
  volume  = {148},
  number  = {2},
  pages   = {778--788},
  year    = {1966},
  doi     = {10.1103/PhysRev.148.778}
}

@article{guo2017heat,
  author  = {Guo, Yangyu and Wang, Moran},
  title   = {Heat Transport in Two-Dimensional Materials by Directly Solving the Phonon Boltzmann Equation under Callaway's Dual Relaxation Model},
  journal = {Physical Review B},
  volume  = {96},
  number  = {13},
  pages   = {134312},
  year    = {2017},
  doi     = {10.1103/PhysRevB.96.134312}
}

@article{bao2018review,
  author  = {Bao, Hua and Chen, Jie and Gu, Xiaokun and Cao, Bing-Yang},
  title   = {A Review of Simulation Methods in Micro/Nanoscale Heat Conduction},
  journal = {ES Energy \& Environment},
  volume  = {1},
  pages   = {16--55},
  year    = {2018},
  doi     = {10.30919/esee8c149}
}

@article{peraud2011efficient,
  author  = {P{\'e}raud, Jean-Philippe M. and Hadjiconstantinou, Nicolas G.},
  title   = {Efficient Simulation of Multidimensional Phonon Transport Using Energy-Based Variance-Reduced Monte Carlo Formulations},
  journal = {Physical Review B},
  volume  = {84},
  pages   = {205331},
  year    = {2011},
  doi     = {10.1103/PhysRevB.84.205331}
}

@article{murthy2002computation,
  author  = {Murthy, Jayathi Y. and Mathur, Sanjay R.},
  title   = {Computation of Sub-Micron Thermal Transport Using an Unstructured Finite Volume Method},
  journal = {Journal of Heat Transfer},
  volume  = {124},
  number  = {6},
  pages   = {1176--1181},
  year    = {2002},
  doi     = {10.1115/1.1518495}
}

@article{mittal2011hybrid,
  title={Hybrid discrete ordinates—spherical harmonics solution to the Boltzmann Transport Equation for phonons for non-equilibrium heat conduction},
  author={Mittal, Arpit and Mazumder, Sandip},
  journal={Journal of Computational Physics},
  volume={230},
  number={18},
  pages={6977--7001},
  year={2011},
  publisher={Elsevier},
  doi = {10.1016/j.jcp.2011.05.024}
}

@article{guo2016dugks,
  author  = {Guo, Zhaoli and Xu, Kun},
  title   = {Discrete Unified Gas Kinetic Scheme for Multiscale Heat Transfer Based on the Phonon Boltzmann Transport Equation},
  journal = {International Journal of Heat and Mass Transfer},
  volume  = {102},
  pages   = {944--958},
  year    = {2016},
  doi     = {10.1016/j.ijheatmasstransfer.2016.06.088}
}

@article{luo2017dugks,
  author  = {Luo, Xiao-Ping and Yi, Hong-Liang},
  title   = {A Discrete Unified Gas Kinetic Scheme for Phonon Boltzmann Transport Equation Accounting for Phonon Dispersion and Polarization},
  journal = {International Journal of Heat and Mass Transfer},
  volume  = {114},
  pages   = {970--980},
  year    = {2017},
  doi     = {10.1016/j.ijheatmasstransfer.2017.06.127}
}

@article{shen2025accurate,
  author  = {Shen, Dingtao and Su, Wei},
  title   = {Accurate-Geometry-Embodied Finite Element Method for Phonon Boltzmann Transport Equation},
  journal = {Computer Physics Communications},
  volume  = {313},
  pages   = {109623},
  year    = {2025},
  doi     = {10.1016/j.cpc.2025.109623}
}

@article{luo2025dg,
  author  = {Luo, Xiao-Ping and Shen, Jia-Wei and Yi, Hong-Liang},
  title   = {Discontinuous Galerkin Finite Element Method for Phonon Boltzmann Transport Equation under Callaway's Scattering Model},
  journal = {Physica E: Low-Dimensional Systems and Nanostructures},
  volume  = {172},
  pages   = {116295},
  year    = {2025},
  doi     = {10.1016/j.physe.2025.116295}
}

@article{adams2002fast,
  author  = {Adams, Marvin L. and Larsen, Edward W.},
  title   = {Fast Iterative Methods for Discrete-Ordinates Particle Transport Calculations},
  journal = {Progress in Nuclear Energy},
  volume  = {40},
  number  = {1},
  pages   = {3--159},
  year    = {2002},
  doi     = {10.1016/S0149-1970(01)00023-3}
}

@article{loy2015coupled,
  author  = {Loy, Jason M. and Mathur, Sanjay R. and Murthy, Jayathi Y.},
  title   = {A Coupled Ordinates Method for Convergence Acceleration of the Phonon Boltzmann Transport Equation},
  journal = {Journal of Heat Transfer},
  volume  = {137},
  number  = {1},
  pages   = {012402},
  year    = {2015},
  doi     = {10.1115/1.4028806}
}

@article{wu2017fast,
  author  = {Wu, Lei and Zhang, Jun and Liu, Haihu and Zhang, Yonghao and Reese, Jason M.},
  title   = {A Fast Iterative Scheme for the Linearized Boltzmann Equation},
  journal = {Journal of Computational Physics},
  volume  = {338},
  pages   = {431--451},
  year    = {2017},
  doi     = {10.1016/j.jcp.2017.03.002}
}

@article{su2020can,
  author  = {Su, Wei and Zhu, Lianhua and Wang, Peng and Zhang, Yonghao and Wu, Lei},
  title   = {Can We Find Steady-State Solutions to Multiscale Rarefied Gas Flows within Dozens of Iterations?},
  journal = {Journal of Computational Physics},
  volume  = {407},
  pages   = {109245},
  year    = {2020},
  doi     = {10.1016/j.jcp.2020.109245}
}

@article{su2020fast,
  author  = {Su, Wei and Zhu, Lianhua and Wu, Lei},
  title   = {Fast Convergence and Asymptotic Preserving of the General Synthetic Iterative Scheme},
  journal = {SIAM Journal on Scientific Computing},
  volume  = {42},
  number  = {6},
  pages   = {B1517--B1540},
  year    = {2020},
  doi     = {10.1137/20M132691X}
}

@article{zhang2021synthetic,
  author  = {Zhang, Chuang and Chen, Songze and Guo, Zhaoli and Wu, Lei},
  title   = {A Fast Synthetic Iterative Scheme for the Stationary Phonon Boltzmann Transport Equation},
  journal = {International Journal of Heat and Mass Transfer},
  volume  = {174},
  pages   = {121308},
  year    = {2021},
  doi     = {10.1016/j.ijheatmasstransfer.2021.121308}
}

@article{zhang2023acceleration,
  author  = {Zhang, Chuang and Huberman, Samuel and Song, Xinliang and Zhao, Jin and Chen, Songze and Wu, Lei},
  title   = {Acceleration Strategy of Source Iteration Method for the Stationary Phonon Boltzmann Transport Equation},
  journal = {International Journal of Heat and Mass Transfer},
  volume  = {217},
  pages   = {124715},
  year    = {2023},
  doi     = {10.1016/j.ijheatmasstransfer.2023.124715}
}

@article{hu2024giftbte,
  author  = {Hu, Yue and Jia, Ru and Xu, Jiaxuan and Sheng, Yufei and Wen, Minhua and Lin, James and Shen, Yongxing and Bao, Hua},
  title   = {{GiftBTE}: An Efficient Deterministic Solver for Non-Gray Phonon Boltzmann Transport Equation},
  journal = {Journal of Physics: Condensed Matter},
  volume  = {36},
  pages   = {025901},
  year    = {2024},
  doi     = {10.1088/1361-648X/acfdea}
}

@article{zhang2025hotspot,
  author  = {Zhang, Chuang and Lou, Qin and Liang, Hong},
  title   = {Synthetic Iterative Scheme for Thermal Applications in Hotspot Systems with Large Temperature Variance},
  journal = {International Journal of Heat and Mass Transfer},
  volume  = {236},
  pages   = {126374},
  year    = {2025},
  doi     = {10.1016/j.ijheatmasstransfer.2024.126374}
}

@misc{chen2026nanobte,
  author        = {Chen, Hongjiang and Lin, Hai-Xuan and Tian, Xiaole and Chen, Hongyu and Zhang, Juan and Yan, Shengyao and Shao, Hezhu and Wu, Yu and Liu, Junjie and Zhang, Hao},
  title         = {{NanoBTE}: Fast Iterative Solution of the Phonon Boltzmann Transport Equation for Nanoscale Heat Transport},
  year          = {2026},
  eprint        = {2607.03226},
  archiveprefix = {arXiv},
  primaryclass  = {cond-mat.mtrl-sci},
  doi           = {10.48550/arXiv.2607.03226},
  note          = {Version 3, 24 July 2026}
}

@article{liu2022fast,
  author  = {Liu, Jia and Zhang, Chuang and Yuan, Haizhuan and Su, Wei and Wu, Lei},
  title   = {A Fast-Converging Scheme for the Phonon Boltzmann Equation with Dual Relaxation Times},
  journal = {Journal of Computational Physics},
  volume  = {467},
  pages   = {111436},
  year    = {2022},
  doi     = {10.1016/j.jcp.2022.111436}
}

@article{luo2025porous,
  author  = {Luo, Xiao-Ping and Guo, Yangyu and Yi, Hong-Liang},
  title   = {Phonon Hydrodynamics in Porous Graphene from Direct Solution of the Boltzmann Equation},
  journal = {Materials Today Physics},
  volume  = {58},
  pages   = {101855},
  year    = {2025},
  doi     = {10.1016/j.mtphys.2025.101855}
}

@article{li2021pinn,
  author  = {Li, Ruiyang and Lee, Eungkyu and Luo, Tengfei},
  title   = {Physics-Informed Neural Networks for Solving Multiscale Mode-Resolved Phonon Boltzmann Transport Equation},
  journal = {Materials Today Physics},
  volume  = {19},
  pages   = {100429},
  year    = {2021},
  doi     = {10.1016/j.mtphys.2021.100429}
}

@article{zhou2023pinn,
  author  = {Zhou, Jiahang and Li, Ruiyang and Luo, Tengfei},
  title   = {Physics-Informed Neural Networks for Solving Time-Dependent Mode-Resolved Phonon Boltzmann Transport Equation},
  journal = {npj Computational Materials},
  volume  = {9},
  pages   = {212},
  year    = {2023},
  doi     = {10.1038/s41524-023-01165-7}
}

@article{shang2025jaxbte,
  author  = {Shang, Wenjie and Zhou, Jiahang and Panda, J. P. and Xu, Zhihao and Liu, Yi and Du, Pan and Wang, Jian-Xun and Luo, Tengfei},
  title   = {{JAX-BTE}: A GPU-Accelerated Differentiable Solver for Phonon Boltzmann Transport Equations},
  journal = {npj Computational Materials},
  volume  = {11},
  pages   = {129},
  year    = {2025},
  doi     = {10.1038/s41524-025-01635-0}
}

@article{wang2022pinn,
  author  = {Wang, Sifan and Yu, Xinling and Perdikaris, Paris},
  title   = {When and Why {PINNs} Fail to Train: A Neural Tangent Kernel Perspective},
  journal = {Journal of Computational Physics},
  volume  = {449},
  pages   = {110768},
  year    = {2022},
  doi     = {10.1016/j.jcp.2021.110768}
}

@article{jin2022asymptotic,
  title={Asymptotic-preserving schemes for multiscale physical problems},
  author={Jin, Shi},
  journal={Acta Numerica},
  volume={31},
  pages={415--489},
  year={2022},
  publisher={Cambridge University Press},
  doi = {10.1017/S0962492922000010}
}

@article{Shang2020HydroPhonon,
  title={Heat vortex in hydrodynamic phonon transport of two-dimensional materials},
  author={Shang, Man-Yu and Zhang, Chuang and Guo, Zhaoli and L\u{u}, Jing-tao},
  journal={Scientific Reports},
  volume={10},
  pages={8272},
  year={2020},
  doi={ttps://doi.org/10.1038/s41598-020-65221-8}
}

@article{Fryer2014moment,
  title={Moment model and boundary conditions for energy transport in the phonon gas},
  author={Fryer, M.J. and Struchtrup, H.},
  journal={Continuum Mechanics and Thermodynamics},
  volume={26},
  pages={593–618},
  year={2014},
  doi={https://doi.org/10.1007/s00161-013-0320-y}
}

@book{chapman1990mathematical,
  title={The Mathematical Theory of Non-uniform Gases: An Account of the Kinetic Theory of Viscosity, Thermal Conduction and Diffusion in Gases},
  author={Chapman, S. and Cowling, T. G.},
  isbn={9780521408448},
  lccn={70077285},
  series={Cambridge Mathematical Library},
  year={1990},
  publisher={Cambridge University Press}
}

@article{Mazumder2022,
  title={{B}OLTZMANN TRANSPORT EQUATION BASED MODELING OF PHONON HEAT CONDUCTION: {P}ROGRESS AND CHALLENGES},
  author={Sandip Mazumder},
  journal={Annual Reviews of Heat Transfer},
  volume={24},
  pages={71-130},
  year={2022},
  hide={https://doi.org/10.1615/AnnualRevHeatTransfer.2022041316}
}

@misc{liu2025ugkwpiugkp,
      title={{UGKWP} and {IUGKP} methods for Multi-Scale Phonon Transport with Dispersion and Polarization}, 
      author={Hongyu Liu and Xiaojian Yang and Chuang Zhang and Xing Ji and Kun Xu},
      year={2025},
      eprint={2506.16203},
      archivePrefix={arXiv},
      primaryClass={physics.comp-ph},
}

@article{liu2025ugkwp,
  title = {Unified gas-kinetic wave-particle method for multiscale phonon transport},
  author = {Liu, Hongyu and Yang, Xiaojian and Zhang, Chuang and Ji, Xing and Xu, Kun},
  journal = {Physical Review E},
  volume = {112},
  pages = {065304},
  numpages = {13},
  year = {2025},
  doi = {10.1103/hz9s-5qbm},
}

@article{Sevilla2018HDG,
  title = {{HDG-NEFEM} with Degree Adaptivity for Stokes Flows},
  author = {Sevilla, R. and Huerta, A.},
  journal = {Journal of Scientific Computing},
  volume = {77},
  pages = {1953–1980},
  year = {2018},
  doi = {10.1007/s10915-018-0657-2},
}

@book{NodalDG,
  title={Nodal Discontinuous Galerkin Methods: Algorithms, Analysis, and Applications},
  author={Jan S. Hesthaven and Tim Warburton},
  year={2008},
  publisher={Springer New York},
  doi= {10.1007/978-0-387-72067-8}
}

\end{document}